\documentclass[%
 reprint,
superscriptaddress,
 amsmath,amssymb,
 aps,
 showkeys,
]{revtex4-2}
\usepackage{soul}
\usepackage{graphicx}
\usepackage{dcolumn}
\usepackage{bm}
\usepackage{hyperref}

\usepackage[version=3]{mhchem} 
\usepackage{siunitx}
\usepackage{lipsum}
\usepackage{physics}
\usepackage{comment}
\usepackage{setspace}

\usepackage{xr}
\usepackage{verbatim}

\begin{document}

\preprint{APS/123-QED}

\title{Low-Frequency Charge Noise in Bilayer Graphene Quantum Dots
}

\author{Jessica Richter}
\email{jrichter@phys.ethz.ch}
\affiliation{%
 Laboratory for Solid State Physics, ETH Zürich, CH-8093 Zürich, Switzerland
}%
\author{Max J. Ruckriegel}
\affiliation{%
 Laboratory for Solid State Physics, ETH Zürich, CH-8093 Zürich, Switzerland
}%
\author{Jonas D. Gerber}
\affiliation{%
 Laboratory for Solid State Physics, ETH Zürich, CH-8093 Zürich, Switzerland
}%
\author{Tijl Degroote}
\affiliation{%
 Laboratory for Solid State Physics, ETH Zürich, CH-8093 Zürich, Switzerland
}%
\author{Christoph Adam}
\affiliation{%
 Laboratory for Solid State Physics, ETH Zürich, CH-8093 Zürich, Switzerland
}%
\author{Markus Niese}
\affiliation{%
 Laboratory for Solid State Physics, ETH Zürich, CH-8093 Zürich, Switzerland
}%
\author{Lara Ostertag}
\affiliation{%
 Laboratory for Solid State Physics, ETH Zürich, CH-8093 Zürich, Switzerland
}%
\author{Clara Galante-Agero}
\affiliation{%
 Laboratory for Solid State Physics, ETH Zürich, CH-8093 Zürich, Switzerland
}%
\author{Kenji Watanabe}
\affiliation{%
 Research Center for Electronic and Optical Materials, National Institute for Materials Science, 1-1 Namiki, Tsukuba 305-0044, Japan
}%
\author{Takashi Taniguchi}
\affiliation{%
 Research Center for Materials Nanoarchitectonics, National Institute for Materials Science,  1-1 Namiki, Tsukuba 305-0044, Japan
}%
\author{Petar Tomi\'c}
\affiliation{%
 Laboratory for Solid State Physics, ETH Zürich, CH-8093 Zürich, Switzerland
}%
\author{Artem O. Denisov}
\affiliation{%
 Laboratory for Solid State Physics, ETH Zürich, CH-8093 Zürich, Switzerland
}%
\author{Hadrien Duprez}
\affiliation{%
 Laboratory for Solid State Physics, ETH Zürich, CH-8093 Zürich, Switzerland
}%
 \author{Klaus Ensslin}%
\author{Thomas Ihn}%
\affiliation{%
 Laboratory for Solid State Physics, ETH Zürich, CH-8093 Zürich, Switzerland
}%
\affiliation{%
 Quantum Center, ETH Zürich, CH-8093 Zürich, Switzerland
}%



\date{\today}

\begin{abstract}
Abstract: Bilayer graphene (BLG) quantum dots (QDs) are a promising platform for semiconductor qubits. However, the low-frequency charge noise that may ultimately limit coherence has remained largely unexplored. Here, we systematically characterize charge noise in gate-defined BLG QDs using transport-based noise spectroscopy. We extract a median amplitude of 
$ S_\mu^{1/2} ( \SI{1}{Hz}) =\SI{1.16}{\micro eV/\sqrt{Hz}}$, placing BLG well within the range reported for established semiconductor quantum-dot platforms. Across variations in charge occupation, confinement, source-drain bias, and charge-sensor operating conditions, neither the noise amplitude nor the spectral dependence shows a reproducible trend in electrostatic tuning, indicating that we extracted the intrinsic semiconductor noise. Consistent noise levels are further observed in double QDs and confirmed using an independent superconducting resonator-based dispersive readout. Extending the study to BLG devices incorporating transition metal dichalcogenide layers reveals no measurable charge noise increase in weakly proximitized QDs. These results validate BLG as a viable platform for coherent quantum information processing.





\end{abstract}


\maketitle


Bilayer graphene (BLG) quantum dots (QDs) combine gate-defined confinement and electrostatic tunability with the ability to engineer spin and valley degrees of freedom while avoiding substantial structural disorder. These complementary properties make BLG a promising platform for semiconductor qubit implementation \cite{tong_pauli_2024, denisov_non-equilibrium_2026, denisov_spinvalley_2025, gachter_single-shot_2022, garreis_long-lived_2024, hecker_coherent_2023}. 
Recent progress has demonstrated single and double quantum dot operation, spin- and valley-control, and integration of tunable spin--orbit coupling \cite{gerber_tunable_2025, dulisch_electric-field-tunable_2025}. To evaluate the suitability of BLG for quantum information processing, benchmarking the relevant decoherence mechanisms is essential.

Among these benchmarks, charge noise is a central figure of merit. Low-frequency electrostatic fluctuations shift the quantum-dot electrochemical potential $\mu$ and couple to the qubit energy via direct-Coulomb, Coulomb-exchange interactions, or spin--orbit coupling, causing qubit dephasing \cite{paladino_1f_2014, burkard_semiconductor_2023}.
Therefore, low-frequency charge noise has become a standard benchmark across established semiconductor platforms.
It is commonly characterized by its power spectral density (PSD) of the form
$ S_\mu(f)= S_1 / f^{\gamma}, $
where $S_1$ denotes the PSD at $\SI{1}{Hz}$ and $\gamma$ the frequency exponent. 
These two fit parameters, $S_1$ and $\gamma$, are used throughout the manuscript to characterize the noise magnitude and spectral shape.

The microscopic origin of charge noise is commonly attributed to ensembles of charge traps switching between two metastable states and other two-level fluctuators (TLFs) located in dielectrics, at interfaces, and in the surrounding material environment \cite{fleetwood_effects_1993, paladino_1f_2014, fleetwood_origins_2020}.  
The typical $1/f$ frequency dependence ($\gamma\approx1$) can be modeled by an ensemble of TLFs uniformly distributed in switching frequency and coupling strength, whereas single TLFs exhibit a Lorentzian spectrum  ($\gamma\ll1$ below and $\gamma\approx2$ above their characteristic switching frequency) \cite{shehata_modeling_2023}. As a result, values of $\gamma$ between 0.2 and 2 are commonly reported \cite{kranz_exploiting_2020}. The charge noise amplitude $S_1$ is a direct measure of the coupling strength and number of TLFs in the surroundings. Typical values span several orders of magnitude and depend on materials and fabrication details \cite{dutta_low-frequency_1981, shehata_modeling_2023}.

In terms of noise sources, bilayer graphene quantum dots represent a largely unexplored system. While conventionally grown semiconductor heterostructures rely on epitaxial growth under ultra-high vacuum conditions to achieve clean interfaces, BLG QDs are realized in van-der-Waals stacks assembled from atomically flat, chemically inert two-dimensional materials. Encapsulation in hexagonal boron nitride (hBN) 
creates an atomically-flat surface that is free of dangling bonds and with negligible defects, as well as only a small lattice constant mismatch with graphene \cite{giovannetti_substrate-induced_2007, zhang_two_2017},
which promises a comparatively clean electrostatic environment \cite{balandin_low-frequency_2013}. However, the BLG-hBN stack is assembled under ambient conditions, using exfoliation and dry-transfer techniques that are more susceptible to contamination than the ultra-high vacuum conditions of a molecular-beam epitaxy machine. Whether these opposing effects ultimately improve or deteriorate charge noise in BLG QDs has remained an open question. 

\begin{figure}[t!]
    \centering
    \includegraphics[width=1\linewidth]{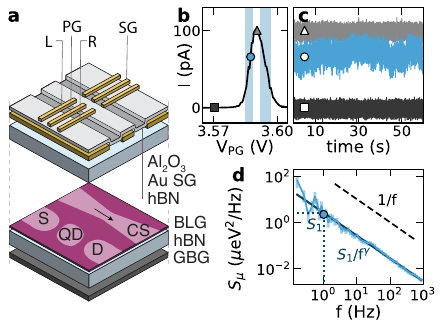}
    \caption{\textbf{Low-frequency charge-noise measurement. }
    \textbf{(a)}  
    Heterostructure that consists of BLG encapsulated in hBN, a graphite back gate (GBG), and two gold top-gate layers. Split gates (SGs) define two parallel transport channels; six finger gates are operated as a plunger gate (PG) and left and right barrier gate (L, R) to define a quantum dot (QD) and a charge sensor (CS), operated with the sensor gate (SG).
    \textbf{(b)} Example of a CB resonance with three operation points, steepest point on flank (blue circle), top of the resonance (gray triangle) and background (black square). Blue-shaded regions mark the flanks. 
    \textbf{(c)} Current fluctuation as a function of time. 
    \textbf{(d)} PSD of the energy fluctuations (light blue operation point) and fit (dark blue). A $1/f$ dependence is also shown for comparison (dashed black line).
    }
    \label{fig:fig_1_method}
\end{figure}

Our measurements are performed on a van-der-Waals  heterostructure consisting of hBN encapsulated BLG, a graphite back gate, and metallic top gates patterned by electron-beam lithography. The metallic top gates consist of two gold layers that are separated by an Al$_2$O$_3$ layer, applied by atomic layer deposition (ALD). 
Figure~\ref{fig:fig_1_method}(a) shows the device and the top gate layout, which enables formation of a single QD controlled by three finger gates (FG) as well as a charge sensor (CS) in parallel to the QD. The device and its fabrication have been previously described in Ref.~\cite{duprez_spin-valley_2024}.

We characterize low-frequency charge noise by measuring transport through a gate-defined quantum dot in the single-level transport regime at a cryogenic temperature of \SI{20}{mK}.
Figures~\ref{fig:fig_1_method}(b-d) illustrate the measurement procedure, with (b) showing a typical Coulomb blockade (CB) resonance as a function of plunger gate voltage $V_\textnormal{PG}$ and (c) reporting current traces $I(t)$ as a function of time at three points of the CB resonance.
At the flank of the CB resonance (blue-shaded region in (b)), small shift in the electrochemical potential $\mu$ of the dot translate most strongly into fluctuations in current (see Supplementary Material S1). 
This flank method is widely used to extract electrochemical-potential fluctuations in semiconductor QDs \cite{dutta_low-frequency_1981,  shehata_modeling_2023, connors_low-frequency_2019, stehouwer_exploiting_2025}.

An alternative benchmark to the charge noise amplitude $S_1$ and the frequency exponent $\gamma$ is the integrated charge noise $\sigma_{\mu}$ \cite{kranz_exploiting_2020, petersson_quantum_2010}, which gives an average noise measure, 
\begin{equation}\label{eq:charge_noise_integration}
    \sigma_\mu = \sqrt{ \int_{f_\text{low}}^{f_\text{high}} S_\mu(f)  ~\mathrm{d}f } .
\end{equation}
For the integration frequency range, we choose the cut-off frequencies that are set by the measurement time $f_\text{low}=1/T_m$ and the sampling frequency $f_\text{high}=f_s.$
This quantity does not require the noise to follow a strict $1/f^\gamma$ spectrum, since it integrates over the measured PSD $S_\mu(f)$, and hence accounts for nonlinearities in the spectrum. This makes $\sigma_\mu$ a robust metric for comparison. Also, the full frequency range spanned by a qubit experiment influences its coherence. So to give a direct indication on BLGs applicability for qubit implementation, we include $\sigma_\mu$ as an additional benchmark. 

In this work, we present a systematic analysis of charge noise in BLG, investigating whether this platform can match the noise characteristics of established QD systems. 
We determine a median noise amplitude, which is competitive with traditional semiconductor platforms, as detailed below. We vary charge occupation, confinement potential, source--drain bias, and charge-sensor operating conditions to evaluate whether these parameters modulate the QDs susceptibility to charge noise or the TLF environment itself. Extending the analysis to a double quantum dot, as well as to resonator-based dispersive readout, spans the study over three different devices and allows us to assess the robustness of the noise characteristics. Moreover, motivated by the prospect of engineering spin--orbit coupling via proximity effects, we examine a transition metal dichalcogenide (TMD) and BLG heterostructure. Lastly, based on our measurement results, we estimate charge qubit relaxation and coherence times, which support BLG as a promising platform for semiconductor qubits.

\begin{figure}[b!]
    \centering
    \includegraphics[width=1\linewidth]{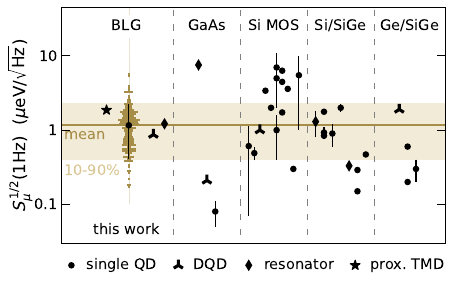}
    \caption{\textbf{Charge-noise amplitude $S_\mu^{1/2}$ at \SI{1}{Hz} of BLG compared to other established QD platforms. }
    Our results for BLG single QDs are summarized in the vertical histogram around the median value (solid yellow line), with the 10-90 percentile range of this distribution highlighted in beige. Additional BLG points reflect varying dot geometries (DQD), readout mechanism (resonator) and heterostructure designs (proximitized TMD).
    Literature values (representative, non-exhaustive) are shown for GaAs 
    \cite{basset_evaluating_2014, dial_charge_2013, jung_background_2004}, Si MOS \cite{elsayed_low_2024, freeman_comparison_2016, jock_silicon_2022, kim_low-disorder_2019, petit_spin_2018, rudolph_long-term_2019, spence_probing_2023, stuyck_uniform_2021, tomic_long_2025,  zwerver_qubits_2022}, Si/SiGe \cite{connors_charge-noise_2022, connors_low-frequency_2019, degli_esposti_low_2024, freeman_comparison_2016, mi_landau-zener_2018, paquelet_wuetz_reducing_2023, struck_low-frequency_2020} and  Ge/SiGe \cite{hendrickx_sweet-spot_2024, lodari_low_2021, stehouwer_exploiting_2025}.
    A full table of the references is provided in the Supplementary Material S2.
    }
    \label{fig:fig_1_comparison}
\end{figure}

Across an ensemble of measurements covering various tuning configurations, we extract a median charge-noise amplitude at $\SI{1}{Hz}$ of
\begin{align*}
    S_\mu^{1/2} (\SI{1}{Hz}) = \sqrt{S_1} = \SI{1.16}{\micro eV/\sqrt{Hz}}.
\end{align*}
We report this value as the representative figure of merit for benchmarking against other material platforms (see S3 for detailed discussion of the statistics). 
Figure~\ref{fig:fig_1_comparison} shows the corresponding histogram of all extracted charge noise amplitudes for the single-dot measurements.
Individual measurements scatter significantly even under nominally similar tuning conditions, a spread whose origin and possible tuning-dependence we analyze in a later section. The 10-90 percentile spread, ranging from 0.41 to \SI{2.24}{\micro eV/\sqrt{Hz}}, is also highlighted in Figure~\ref{fig:fig_1_comparison} (beige shade) along with the median (dark yellow horizontal line).


Charge-noise amplitudes at $\SI{1}{Hz}$ across established semiconductor QD materials including GaAs, Si MOS, Si/SiGe quantum wells and Ge/SiGe devices typically span from $\qtyrange{0.1}{10}{\micro eV/\sqrt{Hz}}$ \cite{burkard_semiconductor_2023}. Our value for $S_1$ places bilayer graphene well within the reported range. A graphical comparison of representative, non-exhaustive literature values is given in Figure~\ref{fig:fig_1_comparison}.

While direct comparison of individual data points should be interpreted with caution due to differences in device architectures and extraction procedures, the overall similarity of charge-noise amplitudes in BLG and the other materials is clear.
The observed agreement indicates that low-frequency electrostatic fluctuations are not elevated in BLG despite its distinct fabrication process and van-der-Waals heterostructure design. 
This finding allows for different interpretations. First, the intrinsically inert bulk properties of BLG and hBN may be sufficient to match the charge stability achieved through ultrahigh-vacuum and epitaxial growth techniques of conventional semiconductor platforms, despite BLG heterostructure assembly taking place under ambient conditions. 
Alternatively, differences between these fabrication procedures might be masked by a common, dominant noise source shared across platforms, for instance, amorphous oxide layers or interfaces to metallic gate electrodes, which are present in the gate stack of most gate-defined quantum dot systems. Distinguishing between the actual charge-noise sources would require further systematic studies. The following parameter study is a first attempt toward this goal.
Regardless of the charge noise origin, our results demonstrate that bilayer graphene is a competitive qubit platform from the perspective of low-frequency charge stability.

\begin{figure*}[t!]
    \centering
    \includegraphics[width=1\linewidth]{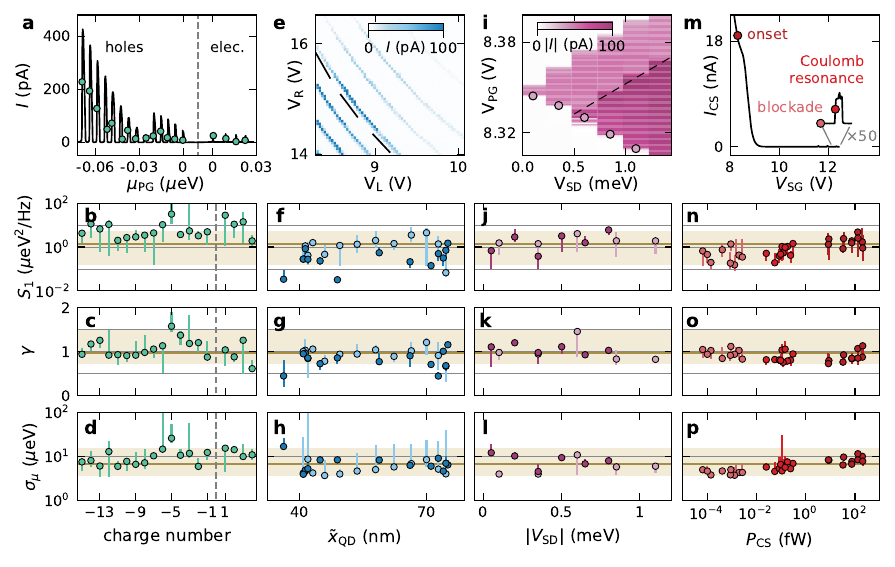}
    \caption{\textbf{Influence of electrostatic tuning on charge noise.}
    The charge-noise amplitude $S_1$ (second row), frequency exponent $\gamma$ (third row) and integrated charge noise $\sigma_\mu$ (fourth row) are shown with respect to four independent tuning parameters. We included the median value, as well as the 10-90 percentile spread of the full measurement ensemble (comp. Sec.~S3). 
    For each charge noise metric, the circle indicates the result of the operation point (at the steepest slope) whereas the vertical bar denotes the range of lowest to highest value measured on the entire flank of the Coulomb resonance (see Supplementary Material S1). 
    \textbf{(a–d) Charge occupation.} In (a) the analyzed Coulomb resonances and their operation points (green dots at steepest slope) are shown as a function of electrochemical potential $\mu$, ranging from 15 holes (negative) to four electrons (positive). 
    \textbf{(e–h) Confinement Tuning.} (e) shows the gate-gate map of several Coulomb resonances in transport. For this, the left (L) and right (R) barrier gate voltages are varied (compare Fig.~\ref{fig:fig_1_method}(a)). Along one Coulomb resonance, the change of relative gate lever arm is converted into an effective position $\tilde{x}_\text{QD}$ (see Sec.~S4). 
    The color difference in (f–h) reflects two repeated measurements along the same Coulomb resonance. 
    \textbf{(i–l) Source-drain bias $|V_\text{SD}|$}. (i) shows the Coulomb resonance as a function of $|V_\text{SD}|$, measurement points are indicated with light pink circles. At $\SI{0.5}{meV}$, an excited state enters into the bias window (dashed line). 
    The two shades of pink used in (j–k) correspond to opposite bias polarity (dark = negative bias, light = positive bias). 
    \textbf{(m–p) Charge-sensor back-action}. (m) shows the sensor current $I_\text{CS}$ as a function of sensor-gate voltage $V_\text{SG}$ (compare Fig.~\ref{fig:fig_1_method}(a)). Three sensor operation regimes are studied: Coulomb blockade (light red), resonance flank (red), and current constriction (dark red).
    Charge noise parameters (n–p) are reported as a function of dissipated power $P_\text{CS}$. 
    }
    \label{fig:fig_2}
\end{figure*}


The spread of the charge noise amplitude across similar quantum dot operation conditions (Fig.~\ref{fig:fig_1_comparison}) raises the question of whether it reflects a systematic dependence on the electrostatic parameters, or whether it originates from stochastic fluctuations in the local charge-trap environment that are inaccessible to tuning. Distinguishing between these two scenarios could provide insight into the microscopic origin of the noise. To this end, we systematically vary four independent tuning parameters (Fig.~\ref{fig:fig_2}): QD charge occupation, confinement potential, source-drain bias, and charge-sensor operating power.
These measurements probe carrier screening, proximity to localized fluctuators, activation of higher-energy transport processes, and measurement back-action, respectively.

We vary the quantum dot occupation across the hole and electron regimes, while maintaining transport through well-defined Coulomb blockade resonances. Neither $S_1$, $\gamma$ nor $\sigma_\mu$ shows systematic evolution with carrier number (Fig.~\ref{fig:fig_2}(b–d)), indicating that increasing occupation of the QD does not measurably change its susceptibility to the surrounding electrostatic fluctuations. This is in contrast to a screening effect that has previously been observed in Si QDs \cite{paquelet_wuetz_reducing_2023}, but could not be confirmed by other Si studies \cite{elsayed_low_2024}. We interpret the measured absence of charge-noise screening at high occupation numbers as an indication that the charge traps are either located in direct vicinity to the QD and with a narrow TLF separation, making screening inefficient on the length scale of the dot diameter, or very far away, also reducing the effect of additional charges.

To probe sensitivity to confinement geometry and TLF proximity, we tune the barrier gates while following a single CB resonance (Fig.~\ref{fig:fig_2}(e)). We extract the change of relative gate lever arm and convert it to an effective shift in dot position $\tilde{x}_\text{QD}$ (see Sec.~S4). 
Although individual measurements vary from each other, no systematic trend  with respect to the effective dot position is observed (Fig.~\ref{fig:fig_2}(f–h)). Moreover, these variations are not reproducible between repeated measurements (light vs dark blue points), indicating that they do not only arise from local differences in the material environment, but instead also from electrostatic changes that occur over time. This is pointing away from a picture of few dominant TLFs in vicinity of the QD and toward an ever-changing, uncorrelated ensemble of fluctuators.

To test whether enhanced relaxation processes activate additional fluctuators, we vary the bias applied between source and drain contact from 0 to $\pm\SI{1}{meV}$.
At large bias, the electrons tunneling through the QD dissipate energy in the contacts.
When transport through an excited state becomes possible, the relaxation can also take place near or inside the QD, leading to enhanced charge noise levels, as observed in Ref.~\cite{jung_background_2004}. 
Despite investigating both of these regimes with the extended transport window and the introduction of the first excited state (dashed line Fig.~\ref{fig:fig_2}(i)), all three quantities $S_1$, $\gamma$ and $\sigma_\mu$ remain unchanged within the statistical uncertainty. They also do not show a dependence on the bias direction (dark vs. light color in Fig.~\ref{fig:fig_2}(j–l)).

Lastly, we investigate whether power dissipation in a nearby charge-sensor contributes to the measured charge noise. The charge-sensor is operated across several conductance regimes, namely in the regime of a confined current channel, in the quantum dot regime and in a fully pinched-off regime, as indicated in Figure~\ref{fig:fig_2}(m). The combination of sensor operation mode and sensor bias allows the dissipated power to be varied over a wide range ($10^{-19}$ to $10^{-13}~\text{W}$). 
Nonetheless, no systematic dependence of the extracted noise parameters is observed (Fig.~\ref{fig:fig_2}(n–p)), indicating negligible sensor-induced activation of the TLF environment.

For both mechanisms, elevated source-drain bias and sensor-induced power dissipation, we do not observe enhancement of the charge noise. This suggests a low coupling strength between the dissipated energy and the TLFs. Assuming that the energy is mostly dissipated via phonons and electron--electron scattering processes in BLG, this finding might suggest that the charge traps are not located primarily in the BLG layer itself, in which the phonons and electrons dominantly spread, but rather in the hBN or oxide layers, or at the interface between them.

In the context of activation of the TLF environment, we also investigated temperature dependence that is generally expected to cause a linear increase of the noise level \cite{dutta_low-frequency_1981}. 
Although the trend is not fully conclusive, we do not find a significant rise in charge noise amplitude, frequency exponent or integrated charge noise between an electron temperature of \SI{20}{mK} and \SI{265}{mK} (see Sec.~S5). 
This also suggests a weak coupling of the charge traps to the electronic temperature.

Across all four electrostatic tuning controls reported above, none of the noise characteristics exhibits a reproducible systematic dependence. Instead, they scatter around a stable mean value, with neither indication of systematic change in charge noise susceptibility nor activation or suppression of the TLF landscape. 
Together, these observations suggest that low-frequency charge noise in BLG quantum dots is governed by stochastic fluctuations of the local environment, largely decoupled from the electrostatic and current operating conditions accessible in this study. 
In practice, this demonstrated robustness allows for device operation across the explored parameter space without compromising on charge-noise performance.

\begin{figure}[th!]
    \centering
    \includegraphics[width=\linewidth]{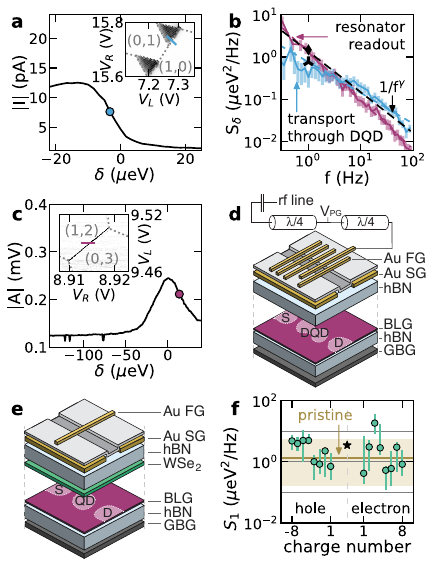}
    \caption{\textbf{Charge noise with respect to quantum dot geometry, readout technique and heterostructure engineering.} 
    \textbf{(a)} Transport signal along the DQD interdot transition (black); operation point at steepest slope (light blue circle, cf. Fig.~\ref{fig:fig_1_method}).
    Inset: charge stability diagram (CSD) with bias triangles and measurement line (light blue); DQD occupation $(N,M)$, with $N$ ($M$) the electron number in the left (right) dot.
    \textbf{(b)} Detuning noise spectrum $S_\delta$ from transport (light blue) and resonator readout (pink), both fitted to $S_1/f^\gamma$  (dashed, blue/pink line). Corresponding \SI{1}{Hz} values refer to Fig.~\ref{fig:fig_1_comparison} (DQD, resonator). Dashed black $S_1/f$ line with median single-dot $S_1$ is shown for comparison.  
    \textbf{(c)} Resonator amplitude $|A|$ (black) across an interdot transition, pink circle marks operation point. Inset: CSD at the (1,2) – (0,3) DQD transition in resonator signal and measurement line (pink).
    \textbf{(d)} Schematic of the resonator device: resonator transmission line coupled to a DQD plunger gate. Abbreviation as in Fig.~\ref{fig:fig_1_method} (a).
    \textbf{(e)} Schematic of the TMD/BLG stack: a WSe$_2$ layer is inserted above the BLG.
    \textbf{(f)} $S_1$ vs. charge occupation for a QD in the TMD/BLG device. Star: TMD median (cf. Fig.~\ref{fig:fig_1_comparison}); yellow line and region: pristine BLG median and 10-90 percentile.  
    }
    \label{fig:fig_3}
\end{figure}

We extend the analysis to double quantum dots (DQDs)  (Fig.~\ref{fig:fig_3}), to test whether the observed charge-noise level generalizes beyond single-dot transport measurements and reflects a more general property of bilayer graphene devices.
In a DQD, the relevant quantity that is affected by charge noise is no longer the electrochemical potential $\mu$  but the level detuning $\delta$ between the two dots. If the charge fluctuations acting on the two dots are only weakly correlated, the detuning noise is expected to remain comparable in amplitude to the single-dot charge noise.

We characterize the detuning noise of a DQD using the same flank-based approach employed for single dots, now utilizing the flank of the baseline of a finite bias triangle across an interdot transition (Fig.~\ref{fig:fig_3}(a)). The resulting charge-noise amplitude $S_1 = \SI{0.76}{\mu eV^2 / Hz}$, frequency exponent $\gamma = 0.73$ and integrated charge noise $\sigma_\delta = \SI{3.53}{\mu eV}$ are consistent with the values obtained in single quantum dots (Fig.~\ref{fig:fig_3}(b) and Fig.~\ref{fig:fig_1_comparison}, DQD). 

The similarity in charge-noise amplitude of the DQD and the single QDs could indicate that the two individual QDs see only weakly correlated noise, and hence stochastically independent TLF environments on the interdot length scale. To confirm this understanding, the spatial correlation of the noise could be measured explicitly, by simultaneous readout of multiple qubits, similar to \cite{tomic_long_2025, yoneda_noise-correlation_2023}, which is beyond the scope of this paper.
This observation nonetheless establishes that introducing an additional confinement region  does not lead to elevated charge noise.

To provide an independent verification of charge noise levels that is not based on transport measurements, we perform charge-noise spectroscopy using a superconducting resonator coupled to a DQD. In this approach, detuning fluctuations in the DQD are inferred from fluctuations of the dispersive resonator response near an interdot transition \cite{basset_evaluating_2014}.

These measurements are performed on a second BLG-hBN heterostructure (previously discussed in Ref.~\cite{ruckriegel_microwave_2026}). The geometry of this device is schematically shown in Figure~\ref{fig:fig_3}(d), where one plunger gate of the DQD is coupled to a superconducting resonator.
The extracted low-frequency noise characteristics ($S_1 = \SI{1.30}{\mu eV^2 / Hz}$, $\gamma = 1.15$, $\sigma_\delta=\SI{2.62}{\mu eV}$) agree well with those obtained from single and DQD transport measurements within experimental uncertainty (Fig.~\ref{fig:fig_3}(b) and Fig.~\ref{fig:fig_1_comparison}, resonator).

The agreement across devices, device geometry, measurement bandwidth, and readout mechanism suggests that the extracted noise levels reflect intrinsic electrostatic fluctuations of the BLG environment rather than artifacts of a particular measurement technique or individual material environment.

Having shown that charge noise in BLG is largely insensitive to QD tuning, device design and readout technique, we also investigated whether this robustness extends to changes in the heterostructure itself. This is particularly relevant for BLG/TMD devices, which are used to induce tunable spin--orbit coupling to BLG QDs \cite{gerber_tunable_2025, dulisch_electric-field-tunable_2025}. 
Introducing the additional material interface, along with defects, vacancies, strain, or lattice mismatch, and adding the proximity to a material with higher intrinsic charge noise \cite{rumyantsev_1_2015, ko_current_2015}, could deteriorate the overall charge-noise level. 

We investigate a BLG/WSe$_2$ heterostructure (Fig.~\ref{fig:fig_3}(e)) \cite{gerber_spin-valley_2026}, in which quantum dots are formed in the weakly proximitized layer, which is the graphene sheet farther away from the WSe$_2$ interface, and hence less strongly affected by the induced spin--orbit coupling.
We compare multiple charge occupations spanning both the electron and hole regimes. Across all measured configurations, the extracted values of $S_1$ (Fig.~\ref{fig:fig_3}(f)), $\gamma$ and $\sigma_\delta$ remain consistent with those obtained in pristine BLG and show no systematic dependence on carrier number or type (see Fig.~\ref{fig:fig_3}(f) and Fig.~\ref{fig:fig_1_comparison}, TMD prox.).

These measurements indicate that introducing a TMD layer does not inherently degrade the low-frequency charge stability of BLG quantum dots. The weakly proximitized BLG layer appears effectively decoupled from additional electrostatic disorder, despite the expected higher intrinsic impurity density in TMDs. 
This effective decoupling of the weakly proximitized BLG layer from the TMD properties is remarkably consistent with previous transport and spectroscopic studies \cite{gerber_tunable_2025, dulisch_electric-field-tunable_2025}.
In total, these results demonstrate that spin--orbit engineering through TMD proximity can be implemented without a measurable charge-noise penalty.


After establishing the charge noise level and its robustness under QD operating conditions and readout mechanisms, we now estimate its significance for qubit implementation in BLG based on the experimental results.
Since electrostatic fluctuations couple directly to the charge degree of freedom, the measured charge noise allows us to predict upper bounds for the relaxation and dephasing times of a double quantum dot charge qubit.

Charge noise affects the relaxation and dephasing of a charge qubit differently: 
the relaxation time is determined by the noise power at the qubit frequency, while dephasing noise accumulates across the full frequency range spanned by the measurement.  
For a representative tunnel coupling of $\Delta/(2h)= \SI{2}{GHz}$ \cite{hecker_coherent_2023}, our median charge-noise level predicts a charge qubit relaxation time at zero detuning of $T_1=\SI{410}{ps}$. 
Dephasing depends on the integrated detuning noise  $\sigma_\delta$, equivalent to the electrochemical potential integrated noise $\sigma_\mu$ defined in Eq. (\ref{eq:charge_noise_integration}) \cite{shnirman_noise_2002, bermeister_charge_2014, petersson_quantum_2010, paquelet_wuetz_reducing_2023, macquarrie_progress_2020, dial_charge_2013}. We capture a typical qubit measurement duration by choosing cut-off frequencies of \SI{1}{mHz} to \SI{5}{GHz} \cite{paquelet_wuetz_reducing_2023, macquarrie_progress_2020}. 
In the quasi-static limit, far away from the sweet spot ($\delta\gg \Delta/2$), the qubit is dephased by fluctuations of the electrochemical potential of each dot, resulting in $T_\varphi= \SI{140}{ps}$. Close to the charge degeneracy point ($\delta\approx0$), the DQD is insensitive to charge noise to first order, typically enhancing $T_\varphi$. Here, we predict a dephasing time of $T_\varphi = \SI{720}{ps}$
\cite{petersson_quantum_2010, basset_evaluating_2014, viennot_out--equilibrium_2014}. 
These estimates extrapolate the charge noise over a wide frequency range, which assumes the $S_1/f^\gamma$ dependence to remain valid over many decades. This has been confirmed in Si QDs \cite{connors_charge-noise_2022, jock_silicon_2022} (Si/SiGe and SiMOS, respectively), but not yet verified for BLG.
Details on the calculation can be found in the Supplementary Material S6. 

Together, the estimated relaxation and dephasing times predict a charge-qubit coherence time of $T_2^* =\SI{390}{ps}$, 
which agrees well with the measured range of $T_2^* = 400-\SI{500}{ps}$ \cite{hecker_coherent_2023}.  
The agreement suggests that at this qubit frequency, coherence is primarily limited by charge noise.
%
This does not seem to be the case at higher charge qubit frequencies. 
A recent study on high frequency noise sources in bilayer graphene \cite{hecker_probing_2026} found a noise spectral density of $\SI{0.5}{neV/\sqrt{Hz}}$ between 4 and $\SI{10}{GHz}$ that was consistent with white-noise characteristics. This allows the conclusion that for a high-frequency charge qubit, electron-phonon coupling and Johnson noise are the main decoherence sources instead of charge noise. This result nicely complements our low-frequency findings and bounds the regime in which our prediction applies.

In the regime of charge noise limited coherence, our estimate for BLG is consistent with $T_2^*$ reported for charge qubits in GaAs quantum dots (\SI{250}{ps} \cite{petersson_quantum_2010}, \SI{400}{ps} \cite{petta_manipulation_2004}), reflecting the comparable charge-noise amplitudes between the two materials (Fig.~\ref{fig:fig_1_comparison}), supporting a direct link of charge-noise level and charge-qubit coherence.

This extrapolation, however, does not carry over to spin and valley qubits, because in BLG the Rashba-type is not the dominant spin--orbit coupling mechanism in the same way as in GaAs, silicon, and germanium. 
Nonetheless, for both spin and valley qubits, charge noise couples only indirectly to the qubit energy through spin--orbit and valley--orbit mixing and should therefore result in significantly longer coherence times than for charge qubits. This is consistent with the measured spin, valley, and spin--valley $T_1$ times, which exceed the estimated charge $T_1$ time by multiple orders of magnitude \cite{banszerus_spin_2022, gachter_single-shot_2022, denisov_spinvalley_2025, banszerus_phonon-limited_2025, garreis_long-lived_2024}. With the strength and character of the spin--orbit and valley--orbit mixing that limit coherence remaining undetermined, we cannot make a quantitative prediction based on the measured charge noise.


To conclude, across this comprehensive study, we established that low-frequency charge noise in BLG QDs, with a median noise amplitude of $S_\mu^{1/2} = \SI{1.16}{\micro eV/\sqrt{Hz}}$ at \SI{1}{Hz}, is comparable to conventional semiconductor platforms, positioning bilayer graphene as a competitive material for qubit applications. 
The charge-noise amplitude, frequency exponent and integrated noise remain largely unaffected across a wide range of different tuning conditions. We observe the absence of a screening effect and no systematic dependence of the charge noise to the quantum dot location. Similarly, we could not report TLF activation, neither by energy dissipated in the contacts or near the charge sensor nor through an increased electron temperature. Even though, this robustness of the charge noise level does not provide an unambiguous understanding of the actual microscopic origin of the charge noise in BLG, it allows us to draw practical conclusions.
We observe that the details of confinement and QD operation do not critically affect charge noise. 
Moreover, we confirm the charge noise level across multiple devices and independent measurement techniques, showing that it reflects an intrinsic charge noise property of BLG-hBN heterostructures. 
Our results also show that TMD proximity engineering can be used to introduce spin--orbit coupling without incurring a significant charge-noise penalty, which is an encouraging outlook for TMD-proximitized BLG as a spin qubit platform.
Together, these findings impose no fundamental constraints on the choice of experimental parameters, device geometry, or heterostructure design, leaving room to target coherence sweet spots.
We find that, from a charge-noise perspective, there are no obstacles for coherent control of spin, valley or spin--valley qubits in BLG.

\begin{acknowledgments}
We are grateful to P. M\"arki, T. B\"ahler, and the FIRST staff for their technical support. We thank Francesco Blanda, Fabrizio Volante and Andrea Hofmann for the valuable collaboration. 
We acknowledge financial support by the Swiss State Secretariat for Education, Research and Innovation SERI, Project number UeM019-7.1, 215929.
\end{acknowledgments}


\newpage
\begin{figure}[h!]
    \centering
    \includegraphics[width=1\linewidth]{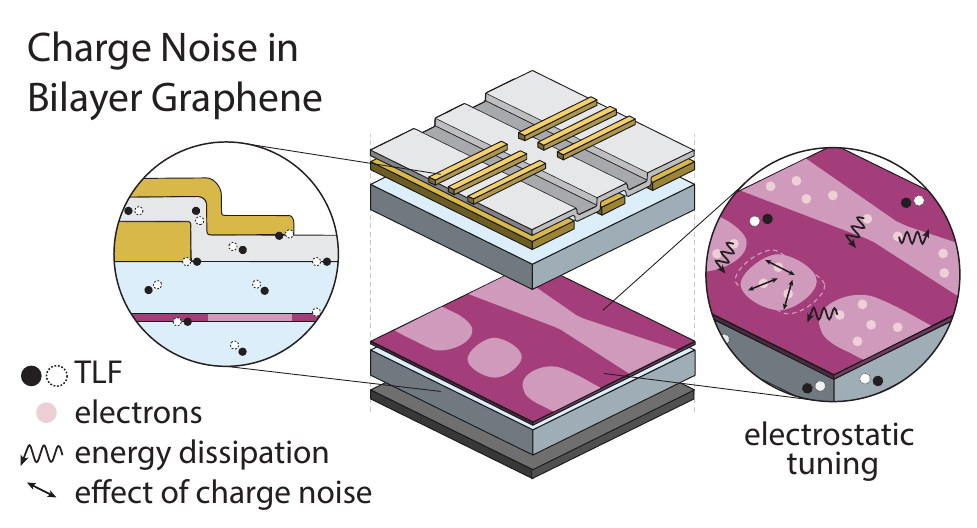}
    \caption{For Table of Contents Only}
\end{figure}



\section*{References}
\bibliography{charge_noise_paper_unicode}

\begin{thebibliography}{55}%
\makeatletter
\providecommand \@ifxundefined [1]{%
 \@ifx{#1\undefined}
}%
\providecommand \@ifnum [1]{%
 \ifnum #1\expandafter \@firstoftwo
 \else \expandafter \@secondoftwo
 \fi
}%
\providecommand \@ifx [1]{%
 \ifx #1\expandafter \@firstoftwo
 \else \expandafter \@secondoftwo
 \fi
}%
\providecommand \natexlab [1]{#1}%
\providecommand \enquote  [1]{``#1''}%
\providecommand \bibnamefont  [1]{#1}%
\providecommand \bibfnamefont [1]{#1}%
\providecommand \citenamefont [1]{#1}%
\providecommand \href@noop [0]{\@secondoftwo}%
\providecommand \href [0]{\begingroup \@sanitize@url \@href}%
\providecommand \@href[1]{\@@startlink{#1}\@@href}%
\providecommand \@@href[1]{\endgroup#1\@@endlink}%
\providecommand \@sanitize@url [0]{\catcode `\\12\catcode `\$12\catcode `\&12\catcode `\#12\catcode `\^12\catcode `\_12\catcode `\%12\relax}%
\providecommand \@@startlink[1]{}%
\providecommand \@@endlink[0]{}%
\providecommand \url  [0]{\begingroup\@sanitize@url \@url }%
\providecommand \@url [1]{\endgroup\@href {#1}{\urlprefix }}%
\providecommand \urlprefix  [0]{URL }%
\providecommand \Eprint [0]{\href }%
\providecommand \doibase [0]{https://doi.org/}%
\providecommand \selectlanguage [0]{\@gobble}%
\providecommand \bibinfo  [0]{\@secondoftwo}%
\providecommand \bibfield  [0]{\@secondoftwo}%
\providecommand \translation [1]{[#1]}%
\providecommand \BibitemOpen [0]{}%
\providecommand \bibitemStop [0]{}%
\providecommand \bibitemNoStop [0]{.\EOS\space}%
\providecommand \EOS [0]{\spacefactor3000\relax}%
\providecommand \BibitemShut  [1]{\csname bibitem#1\endcsname}%
\let\auto@bib@innerbib\@empty
\bibitem [{\citenamefont {Tong}\ \emph {et~al.}()\citenamefont {Tong}, \citenamefont {Kurzmann}, \citenamefont {Garreis}, \citenamefont {Watanabe}, \citenamefont {Taniguchi}, \citenamefont {Ihn},\ and\ \citenamefont {Ensslin}}]{tong_pauli_2024}%
  \BibitemOpen
  \bibfield  {author} {\bibinfo {author} {\bibfnamefont {C.}~\bibnamefont {Tong}}, \bibinfo {author} {\bibfnamefont {A.}~\bibnamefont {Kurzmann}}, \bibinfo {author} {\bibfnamefont {R.}~\bibnamefont {Garreis}}, \bibinfo {author} {\bibfnamefont {K.}~\bibnamefont {Watanabe}}, \bibinfo {author} {\bibfnamefont {T.}~\bibnamefont {Taniguchi}}, \bibinfo {author} {\bibfnamefont {T.}~\bibnamefont {Ihn}},\ and\ \bibinfo {author} {\bibfnamefont {K.}~\bibnamefont {Ensslin}},\ }\bibfield  {title} {\bibinfo {title} {Pauli blockade catalogue and three- and four-particle kondo effect in bilayer graphene quantum dots},\ }\href {https://doi.org/10.1103/PhysRevResearch.6.L012006} {\ \textbf {\bibinfo {volume} {6}},\ \bibinfo {pages} {L012006}}\BibitemShut {NoStop}%
\bibitem [{\citenamefont {Denisov}\ \emph {et~al.}({\natexlab{a}})\citenamefont {Denisov}, \citenamefont {Adam}, \citenamefont {Duprez}, \citenamefont {Richter}, \citenamefont {Chen}, \citenamefont {Hofmann}, \citenamefont {Watanabe}, \citenamefont {Taniguchi}, \citenamefont {Ihn},\ and\ \citenamefont {Ensslin}}]{denisov_non-equilibrium_2026}%
  \BibitemOpen
  \bibfield  {author} {\bibinfo {author} {\bibfnamefont {A.~O.}\ \bibnamefont {Denisov}}, \bibinfo {author} {\bibfnamefont {C.}~\bibnamefont {Adam}}, \bibinfo {author} {\bibfnamefont {H.}~\bibnamefont {Duprez}}, \bibinfo {author} {\bibfnamefont {J.}~\bibnamefont {Richter}}, \bibinfo {author} {\bibfnamefont {Z.}~\bibnamefont {Chen}}, \bibinfo {author} {\bibfnamefont {A.}~\bibnamefont {Hofmann}}, \bibinfo {author} {\bibfnamefont {K.}~\bibnamefont {Watanabe}}, \bibinfo {author} {\bibfnamefont {T.}~\bibnamefont {Taniguchi}}, \bibinfo {author} {\bibfnamefont {T.}~\bibnamefont {Ihn}},\ and\ \bibinfo {author} {\bibfnamefont {K.}~\bibnamefont {Ensslin}},\ }\href {https://doi.org/10.48550/arXiv.2603.00762} {\bibinfo {title} {Non-equilibrium transport reveals energy level degeneracy}} ({\natexlab{a}}),\ \Eprint {https://arxiv.org/abs/2603.00762 [cond-mat.mes-hall]} {2603.00762 [cond-mat.mes-hall]} \BibitemShut {NoStop}%
\bibitem [{\citenamefont {Denisov}\ \emph {et~al.}({\natexlab{b}})\citenamefont {Denisov}, \citenamefont {Reckova}, \citenamefont {Cances}, \citenamefont {Ruckriegel}, \citenamefont {Masseroni}, \citenamefont {Adam}, \citenamefont {Tong}, \citenamefont {Gerber}, \citenamefont {Huang}, \citenamefont {Watanabe}, \citenamefont {Taniguchi}, \citenamefont {Ihn}, \citenamefont {Ensslin},\ and\ \citenamefont {Duprez}}]{denisov_spinvalley_2025}%
  \BibitemOpen
  \bibfield  {author} {\bibinfo {author} {\bibfnamefont {A.~O.}\ \bibnamefont {Denisov}}, \bibinfo {author} {\bibfnamefont {V.}~\bibnamefont {Reckova}}, \bibinfo {author} {\bibfnamefont {S.}~\bibnamefont {Cances}}, \bibinfo {author} {\bibfnamefont {M.~J.}\ \bibnamefont {Ruckriegel}}, \bibinfo {author} {\bibfnamefont {M.}~\bibnamefont {Masseroni}}, \bibinfo {author} {\bibfnamefont {C.}~\bibnamefont {Adam}}, \bibinfo {author} {\bibfnamefont {C.}~\bibnamefont {Tong}}, \bibinfo {author} {\bibfnamefont {J.~D.}\ \bibnamefont {Gerber}}, \bibinfo {author} {\bibfnamefont {W.~W.}\ \bibnamefont {Huang}}, \bibinfo {author} {\bibfnamefont {K.}~\bibnamefont {Watanabe}}, \bibinfo {author} {\bibfnamefont {T.}~\bibnamefont {Taniguchi}}, \bibinfo {author} {\bibfnamefont {T.}~\bibnamefont {Ihn}}, \bibinfo {author} {\bibfnamefont {K.}~\bibnamefont {Ensslin}},\ and\ \bibinfo {author} {\bibfnamefont {H.}~\bibnamefont {Duprez}},\ }\bibfield  {title} {\bibinfo {title} {Spin--valley protected kramers pair in bilayer graphene},\
  }\href {https://doi.org/10.1038/s41565-025-01858-8} {\ \textbf {\bibinfo {volume} {20}},\ \bibinfo {pages} {494} ({\natexlab{b}})}\BibitemShut {NoStop}%
\bibitem [{\citenamefont {G{\"a}chter}\ \emph {et~al.}()\citenamefont {G{\"a}chter}, \citenamefont {Garreis}, \citenamefont {Gerber}, \citenamefont {Ruckriegel}, \citenamefont {Tong}, \citenamefont {Kratochwil}, \citenamefont {de~Vries}, \citenamefont {Kurzmann}, \citenamefont {Watanabe}, \citenamefont {Taniguchi}, \citenamefont {Ihn}, \citenamefont {Ensslin},\ and\ \citenamefont {Huang}}]{gachter_single-shot_2022}%
  \BibitemOpen
  \bibfield  {author} {\bibinfo {author} {\bibfnamefont {L.~M.}\ \bibnamefont {G{\"a}chter}}, \bibinfo {author} {\bibfnamefont {R.}~\bibnamefont {Garreis}}, \bibinfo {author} {\bibfnamefont {J.~D.}\ \bibnamefont {Gerber}}, \bibinfo {author} {\bibfnamefont {M.~J.}\ \bibnamefont {Ruckriegel}}, \bibinfo {author} {\bibfnamefont {C.}~\bibnamefont {Tong}}, \bibinfo {author} {\bibfnamefont {B.}~\bibnamefont {Kratochwil}}, \bibinfo {author} {\bibfnamefont {F.~K.}\ \bibnamefont {de~Vries}}, \bibinfo {author} {\bibfnamefont {A.}~\bibnamefont {Kurzmann}}, \bibinfo {author} {\bibfnamefont {K.}~\bibnamefont {Watanabe}}, \bibinfo {author} {\bibfnamefont {T.}~\bibnamefont {Taniguchi}}, \bibinfo {author} {\bibfnamefont {T.}~\bibnamefont {Ihn}}, \bibinfo {author} {\bibfnamefont {K.}~\bibnamefont {Ensslin}},\ and\ \bibinfo {author} {\bibfnamefont {W.~W.}\ \bibnamefont {Huang}},\ }\bibfield  {title} {\bibinfo {title} {Single-shot spin readout in graphene quantum dots},\ }\href {https://doi.org/10.1103/PRXQuantum.3.020343} {\
  \textbf {\bibinfo {volume} {3}},\ \bibinfo {pages} {020343}}\BibitemShut {NoStop}%
\bibitem [{\citenamefont {Garreis}\ \emph {et~al.}()\citenamefont {Garreis}, \citenamefont {Tong}, \citenamefont {Terle}, \citenamefont {Ruckriegel}, \citenamefont {Gerber}, \citenamefont {G{\"a}chter}, \citenamefont {Watanabe}, \citenamefont {Taniguchi}, \citenamefont {Ihn}, \citenamefont {Ensslin},\ and\ \citenamefont {Huang}}]{garreis_long-lived_2024}%
  \BibitemOpen
  \bibfield  {author} {\bibinfo {author} {\bibfnamefont {R.}~\bibnamefont {Garreis}}, \bibinfo {author} {\bibfnamefont {C.}~\bibnamefont {Tong}}, \bibinfo {author} {\bibfnamefont {J.}~\bibnamefont {Terle}}, \bibinfo {author} {\bibfnamefont {M.~J.}\ \bibnamefont {Ruckriegel}}, \bibinfo {author} {\bibfnamefont {J.~D.}\ \bibnamefont {Gerber}}, \bibinfo {author} {\bibfnamefont {L.~M.}\ \bibnamefont {G{\"a}chter}}, \bibinfo {author} {\bibfnamefont {K.}~\bibnamefont {Watanabe}}, \bibinfo {author} {\bibfnamefont {T.}~\bibnamefont {Taniguchi}}, \bibinfo {author} {\bibfnamefont {T.}~\bibnamefont {Ihn}}, \bibinfo {author} {\bibfnamefont {K.}~\bibnamefont {Ensslin}},\ and\ \bibinfo {author} {\bibfnamefont {W.~W.}\ \bibnamefont {Huang}},\ }\bibfield  {title} {\bibinfo {title} {Long-lived valley states in bilayer graphene quantum dots},\ }\href {https://doi.org/10.1038/s41567-023-02334-7} {\ \textbf {\bibinfo {volume} {20}},\ \bibinfo {pages} {428}}\BibitemShut {NoStop}%
\bibitem [{\citenamefont {Hecker}\ \emph {et~al.}({\natexlab{a}})\citenamefont {Hecker}, \citenamefont {Banszerus}, \citenamefont {Sch{\"a}pers}, \citenamefont {M{\"o}ller}, \citenamefont {Peters}, \citenamefont {Icking}, \citenamefont {Watanabe}, \citenamefont {Taniguchi}, \citenamefont {Volk},\ and\ \citenamefont {Stampfer}}]{hecker_coherent_2023}%
  \BibitemOpen
  \bibfield  {author} {\bibinfo {author} {\bibfnamefont {K.}~\bibnamefont {Hecker}}, \bibinfo {author} {\bibfnamefont {L.}~\bibnamefont {Banszerus}}, \bibinfo {author} {\bibfnamefont {A.}~\bibnamefont {Sch{\"a}pers}}, \bibinfo {author} {\bibfnamefont {S.}~\bibnamefont {M{\"o}ller}}, \bibinfo {author} {\bibfnamefont {A.}~\bibnamefont {Peters}}, \bibinfo {author} {\bibfnamefont {E.}~\bibnamefont {Icking}}, \bibinfo {author} {\bibfnamefont {K.}~\bibnamefont {Watanabe}}, \bibinfo {author} {\bibfnamefont {T.}~\bibnamefont {Taniguchi}}, \bibinfo {author} {\bibfnamefont {C.}~\bibnamefont {Volk}},\ and\ \bibinfo {author} {\bibfnamefont {C.}~\bibnamefont {Stampfer}},\ }\bibfield  {title} {\bibinfo {title} {Coherent charge oscillations in a bilayer graphene double quantum dot},\ }\href {https://doi.org/10.1038/s41467-023-43541-3} {\ \textbf {\bibinfo {volume} {14}},\ \bibinfo {pages} {7911} ({\natexlab{a}})}\BibitemShut {NoStop}%
\bibitem [{\citenamefont {Gerber}\ \emph {et~al.}({\natexlab{a}})\citenamefont {Gerber}, \citenamefont {Ersoy}, \citenamefont {Masseroni}, \citenamefont {Niese}, \citenamefont {Laumer}, \citenamefont {Denisov}, \citenamefont {Duprez}, \citenamefont {Huang}, \citenamefont {Adam}, \citenamefont {Ostertag}, \citenamefont {Tong}, \citenamefont {Taniguchi}, \citenamefont {Watanabe}, \citenamefont {Fal'ko}, \citenamefont {Ihn}, \citenamefont {Ensslin},\ and\ \citenamefont {Knothe}}]{gerber_tunable_2025}%
  \BibitemOpen
  \bibfield  {author} {\bibinfo {author} {\bibfnamefont {J.~D.}\ \bibnamefont {Gerber}}, \bibinfo {author} {\bibfnamefont {E.}~\bibnamefont {Ersoy}}, \bibinfo {author} {\bibfnamefont {M.}~\bibnamefont {Masseroni}}, \bibinfo {author} {\bibfnamefont {M.}~\bibnamefont {Niese}}, \bibinfo {author} {\bibfnamefont {M.}~\bibnamefont {Laumer}}, \bibinfo {author} {\bibfnamefont {A.~O.}\ \bibnamefont {Denisov}}, \bibinfo {author} {\bibfnamefont {H.}~\bibnamefont {Duprez}}, \bibinfo {author} {\bibfnamefont {W.~W.}\ \bibnamefont {Huang}}, \bibinfo {author} {\bibfnamefont {C.}~\bibnamefont {Adam}}, \bibinfo {author} {\bibfnamefont {L.}~\bibnamefont {Ostertag}}, \bibinfo {author} {\bibfnamefont {C.}~\bibnamefont {Tong}}, \bibinfo {author} {\bibfnamefont {T.}~\bibnamefont {Taniguchi}}, \bibinfo {author} {\bibfnamefont {K.}~\bibnamefont {Watanabe}}, \bibinfo {author} {\bibfnamefont {V.~I.}\ \bibnamefont {Fal'ko}}, \bibinfo {author} {\bibfnamefont {T.}~\bibnamefont {Ihn}}, \bibinfo {author} {\bibfnamefont {K.}~\bibnamefont
  {Ensslin}},\ and\ \bibinfo {author} {\bibfnamefont {A.}~\bibnamefont {Knothe}},\ }\bibfield  {title} {\bibinfo {title} {Tunable spin--orbit splitting in bilayer graphene/{WSe}2 quantum devices},\ }\href {https://doi.org/10.1021/acs.nanolett.5c02309} {\ \textbf {\bibinfo {volume} {25}},\ \bibinfo {pages} {12480} ({\natexlab{a}})}\BibitemShut {NoStop}%
\bibitem [{\citenamefont {Dulisch}\ \emph {et~al.}()\citenamefont {Dulisch}, \citenamefont {Emmerich}, \citenamefont {Icking}, \citenamefont {Hecker}, \citenamefont {M{\"o}ller}, \citenamefont {M{\"u}ller}, \citenamefont {Watanabe}, \citenamefont {Taniguchi}, \citenamefont {Volk},\ and\ \citenamefont {Stampfer}}]{dulisch_electric-field-tunable_2025}%
  \BibitemOpen
  \bibfield  {author} {\bibinfo {author} {\bibfnamefont {H.}~\bibnamefont {Dulisch}}, \bibinfo {author} {\bibfnamefont {D.}~\bibnamefont {Emmerich}}, \bibinfo {author} {\bibfnamefont {E.}~\bibnamefont {Icking}}, \bibinfo {author} {\bibfnamefont {K.}~\bibnamefont {Hecker}}, \bibinfo {author} {\bibfnamefont {S.}~\bibnamefont {M{\"o}ller}}, \bibinfo {author} {\bibfnamefont {L.}~\bibnamefont {M{\"u}ller}}, \bibinfo {author} {\bibfnamefont {K.}~\bibnamefont {Watanabe}}, \bibinfo {author} {\bibfnamefont {T.}~\bibnamefont {Taniguchi}}, \bibinfo {author} {\bibfnamefont {C.}~\bibnamefont {Volk}},\ and\ \bibinfo {author} {\bibfnamefont {C.}~\bibnamefont {Stampfer}},\ }\bibfield  {title} {\bibinfo {title} {Electric-field-tunable spin--orbit gap in a bilayer graphene/{WSe}2 quantum dot},\ }\href {https://doi.org/10.1021/acs.nanolett.5c02229} {\ \textbf {\bibinfo {volume} {25}},\ \bibinfo {pages} {10549}}\BibitemShut {NoStop}%
\bibitem [{\citenamefont {Paladino}()}]{paladino_1f_2014}%
  \BibitemOpen
  \bibfield  {author} {\bibinfo {author} {\bibfnamefont {E.}~\bibnamefont {Paladino}},\ }\bibfield  {title} {\bibinfo {title} {1/f noise: Implications for solid-state quantum information},\ }\href {https://doi.org/10.1103/RevModPhys.86.361} {\ \textbf {\bibinfo {volume} {86}},\ \bibinfo {pages} {361}}\BibitemShut {NoStop}%
\bibitem [{\citenamefont {Burkard}\ \emph {et~al.}()\citenamefont {Burkard}, \citenamefont {Ladd}, \citenamefont {Pan}, \citenamefont {Nichol},\ and\ \citenamefont {Petta}}]{burkard_semiconductor_2023}%
  \BibitemOpen
  \bibfield  {author} {\bibinfo {author} {\bibfnamefont {G.}~\bibnamefont {Burkard}}, \bibinfo {author} {\bibfnamefont {T.~D.}\ \bibnamefont {Ladd}}, \bibinfo {author} {\bibfnamefont {A.}~\bibnamefont {Pan}}, \bibinfo {author} {\bibfnamefont {J.~M.}\ \bibnamefont {Nichol}},\ and\ \bibinfo {author} {\bibfnamefont {J.~R.}\ \bibnamefont {Petta}},\ }\bibfield  {title} {\bibinfo {title} {Semiconductor spin qubits},\ }\href {https://doi.org/10.1103/RevModPhys.95.025003} {\ \textbf {\bibinfo {volume} {95}},\ \bibinfo {pages} {025003}}\BibitemShut {NoStop}%
\bibitem [{\citenamefont {Fleetwood}\ \emph {et~al.}()\citenamefont {Fleetwood}, \citenamefont {Winokur}, \citenamefont {Reber}, \citenamefont {Meisenheimer}, \citenamefont {Schwank}, \citenamefont {Shaneyfelt},\ and\ \citenamefont {Riewe}}]{fleetwood_effects_1993}%
  \BibitemOpen
  \bibfield  {author} {\bibinfo {author} {\bibfnamefont {D.~M.}\ \bibnamefont {Fleetwood}}, \bibinfo {author} {\bibfnamefont {P.~S.}\ \bibnamefont {Winokur}}, \bibinfo {author} {\bibfnamefont {R.~A.}\ \bibnamefont {Reber}, \bibfnamefont {Jr.}}, \bibinfo {author} {\bibfnamefont {T.~L.}\ \bibnamefont {Meisenheimer}}, \bibinfo {author} {\bibfnamefont {J.~R.}\ \bibnamefont {Schwank}}, \bibinfo {author} {\bibfnamefont {M.~R.}\ \bibnamefont {Shaneyfelt}},\ and\ \bibinfo {author} {\bibfnamefont {L.~C.}\ \bibnamefont {Riewe}},\ }\bibfield  {title} {\bibinfo {title} {Effects of oxide traps, interface traps, and ``border traps'' on metal-oxide-semiconductor devices},\ }\href {https://doi.org/10.1063/1.353777} {\ \textbf {\bibinfo {volume} {73}},\ \bibinfo {pages} {5058}}\BibitemShut {NoStop}%
\bibitem [{\citenamefont {Fleetwood}()}]{fleetwood_origins_2020}%
  \BibitemOpen
  \bibfield  {author} {\bibinfo {author} {\bibfnamefont {D.~M.}\ \bibnamefont {Fleetwood}},\ }\bibfield  {title} {\bibinfo {title} {Origins of 1/f noise in electronic materials and devices: A historical perspective},\ }in\ \href {https://doi.org/10.1007/978-3-030-37500-3_1} {\emph {\bibinfo {booktitle} {Noise in Nanoscale Semiconductor Devices}}},\ \bibinfo {editor} {edited by\ \bibinfo {editor} {\bibfnamefont {T.}~\bibnamefont {Grasser}}}\ (\bibinfo  {publisher} {Springer International Publishing})\ pp.\ \bibinfo {pages} {1--31}\BibitemShut {NoStop}%
\bibitem [{\citenamefont {Shehata}\ \emph {et~al.}()\citenamefont {Shehata}, \citenamefont {Simion}, \citenamefont {Li}, \citenamefont {Mohiyaddin}, \citenamefont {Wan}, \citenamefont {Mongillo}, \citenamefont {Govoreanu}, \citenamefont {Radu}, \citenamefont {De~Greve},\ and\ \citenamefont {Van~Dorpe}}]{shehata_modeling_2023}%
  \BibitemOpen
  \bibfield  {author} {\bibinfo {author} {\bibfnamefont {M.~M. E.~K.}\ \bibnamefont {Shehata}}, \bibinfo {author} {\bibfnamefont {G.}~\bibnamefont {Simion}}, \bibinfo {author} {\bibfnamefont {R.}~\bibnamefont {Li}}, \bibinfo {author} {\bibfnamefont {F.~A.}\ \bibnamefont {Mohiyaddin}}, \bibinfo {author} {\bibfnamefont {D.}~\bibnamefont {Wan}}, \bibinfo {author} {\bibfnamefont {M.}~\bibnamefont {Mongillo}}, \bibinfo {author} {\bibfnamefont {B.}~\bibnamefont {Govoreanu}}, \bibinfo {author} {\bibfnamefont {I.}~\bibnamefont {Radu}}, \bibinfo {author} {\bibfnamefont {K.}~\bibnamefont {De~Greve}},\ and\ \bibinfo {author} {\bibfnamefont {P.}~\bibnamefont {Van~Dorpe}},\ }\bibfield  {title} {\bibinfo {title} {Modeling semiconductor spin qubits and their charge noise environment for quantum gate fidelity estimation},\ }\href {https://doi.org/10.1103/PhysRevB.108.045305} {\ \textbf {\bibinfo {volume} {108}},\ \bibinfo {pages} {045305}}\BibitemShut {NoStop}%
\bibitem [{\citenamefont {Kranz}\ \emph {et~al.}()\citenamefont {Kranz}, \citenamefont {Gorman}, \citenamefont {Thorgrimsson}, \citenamefont {He}, \citenamefont {Keith}, \citenamefont {Keizer},\ and\ \citenamefont {Simmons}}]{kranz_exploiting_2020}%
  \BibitemOpen
  \bibfield  {author} {\bibinfo {author} {\bibfnamefont {L.}~\bibnamefont {Kranz}}, \bibinfo {author} {\bibfnamefont {S.~K.}\ \bibnamefont {Gorman}}, \bibinfo {author} {\bibfnamefont {B.}~\bibnamefont {Thorgrimsson}}, \bibinfo {author} {\bibfnamefont {Y.}~\bibnamefont {He}}, \bibinfo {author} {\bibfnamefont {D.}~\bibnamefont {Keith}}, \bibinfo {author} {\bibfnamefont {J.~G.}\ \bibnamefont {Keizer}},\ and\ \bibinfo {author} {\bibfnamefont {M.~Y.}\ \bibnamefont {Simmons}},\ }\bibfield  {title} {\bibinfo {title} {Exploiting a single-crystal environment to minimize the charge noise on qubits in silicon},\ }\href {https://doi.org/10.1002/adma.202003361} {\ \textbf {\bibinfo {volume} {32}},\ \bibinfo {pages} {2003361}}\BibitemShut {NoStop}%
\bibitem [{\citenamefont {Dutta}()}]{dutta_low-frequency_1981}%
  \BibitemOpen
  \bibfield  {author} {\bibinfo {author} {\bibfnamefont {P.}~\bibnamefont {Dutta}},\ }\bibfield  {title} {\bibinfo {title} {Low-frequency fluctuations in solids: 1/f noise},\ }\href {https://doi.org/10.1103/RevModPhys.53.497} {\ \textbf {\bibinfo {volume} {53}},\ \bibinfo {pages} {497}}\BibitemShut {NoStop}%
\bibitem [{\citenamefont {Giovannetti}\ \emph {et~al.}()\citenamefont {Giovannetti}, \citenamefont {Khomyakov}, \citenamefont {Brocks}, \citenamefont {Kelly},\ and\ \citenamefont {van~den Brink}}]{giovannetti_substrate-induced_2007}%
  \BibitemOpen
  \bibfield  {author} {\bibinfo {author} {\bibfnamefont {G.}~\bibnamefont {Giovannetti}}, \bibinfo {author} {\bibfnamefont {P.~A.}\ \bibnamefont {Khomyakov}}, \bibinfo {author} {\bibfnamefont {G.}~\bibnamefont {Brocks}}, \bibinfo {author} {\bibfnamefont {P.~J.}\ \bibnamefont {Kelly}},\ and\ \bibinfo {author} {\bibfnamefont {J.}~\bibnamefont {van~den Brink}},\ }\bibfield  {title} {\bibinfo {title} {Substrate-induced band gap in graphene on hexagonal boron nitride: Ab initio density functional calculations},\ }\href {https://doi.org/10.1103/PhysRevB.76.073103} {\ \textbf {\bibinfo {volume} {76}},\ \bibinfo {pages} {073103}}\BibitemShut {NoStop}%
\bibitem [{\citenamefont {Zhang}\ \emph {et~al.}()\citenamefont {Zhang}, \citenamefont {Feng}, \citenamefont {Wang}, \citenamefont {Yang},\ and\ \citenamefont {Wang}}]{zhang_two_2017}%
  \BibitemOpen
  \bibfield  {author} {\bibinfo {author} {\bibfnamefont {K.}~\bibnamefont {Zhang}}, \bibinfo {author} {\bibfnamefont {Y.}~\bibnamefont {Feng}}, \bibinfo {author} {\bibfnamefont {F.}~\bibnamefont {Wang}}, \bibinfo {author} {\bibfnamefont {Z.}~\bibnamefont {Yang}},\ and\ \bibinfo {author} {\bibfnamefont {J.}~\bibnamefont {Wang}},\ }\bibfield  {title} {\bibinfo {title} {Two dimensional hexagonal boron nitride (2d-{hBN}): synthesis, properties and applications},\ }\href {https://doi.org/10.1039/c7tc04300g} {\ \textbf {\bibinfo {volume} {5}},\ \bibinfo {pages} {11992}}\BibitemShut {NoStop}%
\bibitem [{\citenamefont {Balandin}()}]{balandin_low-frequency_2013}%
  \BibitemOpen
  \bibfield  {author} {\bibinfo {author} {\bibfnamefont {A.~A.}\ \bibnamefont {Balandin}},\ }\bibfield  {title} {\bibinfo {title} {Low-frequency 1/f noise in graphene devices},\ }\href {https://doi.org/10.1038/nnano.2013.144} {\ \textbf {\bibinfo {volume} {8}},\ \bibinfo {pages} {549}}\BibitemShut {NoStop}%
\bibitem [{\citenamefont {Duprez}\ \emph {et~al.}()\citenamefont {Duprez}, \citenamefont {Cances}, \citenamefont {Omahen}, \citenamefont {Masseroni}, \citenamefont {Ruckriegel}, \citenamefont {Adam}, \citenamefont {Tong}, \citenamefont {Garreis}, \citenamefont {Gerber}, \citenamefont {Huang}, \citenamefont {G{\"a}chter}, \citenamefont {Watanabe}, \citenamefont {Taniguchi}, \citenamefont {Ihn},\ and\ \citenamefont {Ensslin}}]{duprez_spin-valley_2024}%
  \BibitemOpen
  \bibfield  {author} {\bibinfo {author} {\bibfnamefont {H.}~\bibnamefont {Duprez}}, \bibinfo {author} {\bibfnamefont {S.}~\bibnamefont {Cances}}, \bibinfo {author} {\bibfnamefont {A.}~\bibnamefont {Omahen}}, \bibinfo {author} {\bibfnamefont {M.}~\bibnamefont {Masseroni}}, \bibinfo {author} {\bibfnamefont {M.~J.}\ \bibnamefont {Ruckriegel}}, \bibinfo {author} {\bibfnamefont {C.}~\bibnamefont {Adam}}, \bibinfo {author} {\bibfnamefont {C.}~\bibnamefont {Tong}}, \bibinfo {author} {\bibfnamefont {R.}~\bibnamefont {Garreis}}, \bibinfo {author} {\bibfnamefont {J.~D.}\ \bibnamefont {Gerber}}, \bibinfo {author} {\bibfnamefont {W.}~\bibnamefont {Huang}}, \bibinfo {author} {\bibfnamefont {L.}~\bibnamefont {G{\"a}chter}}, \bibinfo {author} {\bibfnamefont {K.}~\bibnamefont {Watanabe}}, \bibinfo {author} {\bibfnamefont {T.}~\bibnamefont {Taniguchi}}, \bibinfo {author} {\bibfnamefont {T.}~\bibnamefont {Ihn}},\ and\ \bibinfo {author} {\bibfnamefont {K.}~\bibnamefont {Ensslin}},\ }\bibfield  {title} {\bibinfo {title}
  {Spin-valley locked excited states spectroscopy in a one-particle bilayer graphene quantum dot},\ }\href {https://doi.org/10.1038/s41467-024-54121-4} {\ \textbf {\bibinfo {volume} {15}},\ \bibinfo {pages} {9717}}\BibitemShut {NoStop}%
\bibitem [{\citenamefont {Connors}\ \emph {et~al.}({\natexlab{a}})\citenamefont {Connors}, \citenamefont {Nelson}, \citenamefont {Qiao}, \citenamefont {Edge},\ and\ \citenamefont {Nichol}}]{connors_low-frequency_2019}%
  \BibitemOpen
  \bibfield  {author} {\bibinfo {author} {\bibfnamefont {E.~J.}\ \bibnamefont {Connors}}, \bibinfo {author} {\bibfnamefont {J.}~\bibnamefont {Nelson}}, \bibinfo {author} {\bibfnamefont {H.}~\bibnamefont {Qiao}}, \bibinfo {author} {\bibfnamefont {L.~F.}\ \bibnamefont {Edge}},\ and\ \bibinfo {author} {\bibfnamefont {J.~M.}\ \bibnamefont {Nichol}},\ }\bibfield  {title} {\bibinfo {title} {Low-frequency charge noise in si/{SiGe} quantum dots},\ }\href {https://doi.org/10.1103/PhysRevB.100.165305} {\ \textbf {\bibinfo {volume} {100}},\ \bibinfo {pages} {165305} ({\natexlab{a}})}\BibitemShut {NoStop}%
\bibitem [{\citenamefont {Stehouwer}\ \emph {et~al.}()\citenamefont {Stehouwer}, \citenamefont {Yu}, \citenamefont {van Straaten}, \citenamefont {Tosato}, \citenamefont {John}, \citenamefont {Degli~Esposti}, \citenamefont {Elsayed}, \citenamefont {Costa}, \citenamefont {Oosterhout}, \citenamefont {Hendrickx}, \citenamefont {Veldhorst}, \citenamefont {Borsoi},\ and\ \citenamefont {Scappucci}}]{stehouwer_exploiting_2025}%
  \BibitemOpen
  \bibfield  {author} {\bibinfo {author} {\bibfnamefont {L.~E.~A.}\ \bibnamefont {Stehouwer}}, \bibinfo {author} {\bibfnamefont {C.~X.}\ \bibnamefont {Yu}}, \bibinfo {author} {\bibfnamefont {B.}~\bibnamefont {van Straaten}}, \bibinfo {author} {\bibfnamefont {A.}~\bibnamefont {Tosato}}, \bibinfo {author} {\bibfnamefont {V.}~\bibnamefont {John}}, \bibinfo {author} {\bibfnamefont {D.}~\bibnamefont {Degli~Esposti}}, \bibinfo {author} {\bibfnamefont {A.}~\bibnamefont {Elsayed}}, \bibinfo {author} {\bibfnamefont {D.}~\bibnamefont {Costa}}, \bibinfo {author} {\bibfnamefont {S.~D.}\ \bibnamefont {Oosterhout}}, \bibinfo {author} {\bibfnamefont {N.~W.}\ \bibnamefont {Hendrickx}}, \bibinfo {author} {\bibfnamefont {M.}~\bibnamefont {Veldhorst}}, \bibinfo {author} {\bibfnamefont {F.}~\bibnamefont {Borsoi}},\ and\ \bibinfo {author} {\bibfnamefont {G.}~\bibnamefont {Scappucci}},\ }\bibfield  {title} {\bibinfo {title} {Exploiting strained epitaxial germanium for scaling low-noise spin qubits at the micrometre scale},\ }\href
  {https://doi.org/10.1038/s41563-025-02276-w} {\ ,\ \bibinfo {pages} {1}}\BibitemShut {NoStop}%
\bibitem [{\citenamefont {Petersson}\ \emph {et~al.}()\citenamefont {Petersson}, \citenamefont {Petta}, \citenamefont {Lu},\ and\ \citenamefont {Gossard}}]{petersson_quantum_2010}%
  \BibitemOpen
  \bibfield  {author} {\bibinfo {author} {\bibfnamefont {K.~D.}\ \bibnamefont {Petersson}}, \bibinfo {author} {\bibfnamefont {J.~R.}\ \bibnamefont {Petta}}, \bibinfo {author} {\bibfnamefont {H.}~\bibnamefont {Lu}},\ and\ \bibinfo {author} {\bibfnamefont {A.~C.}\ \bibnamefont {Gossard}},\ }\bibfield  {title} {\bibinfo {title} {Quantum coherence in a one-electron semiconductor charge qubit},\ }\href {https://doi.org/10.1103/PhysRevLett.105.246804} {\ \textbf {\bibinfo {volume} {105}},\ \bibinfo {pages} {246804}}\BibitemShut {NoStop}%
\bibitem [{\citenamefont {Basset}\ \emph {et~al.}()\citenamefont {Basset}, \citenamefont {Stockklauser}, \citenamefont {Jarausch}, \citenamefont {Frey}, \citenamefont {Reichl}, \citenamefont {Wegscheider}, \citenamefont {Wallraff}, \citenamefont {Ensslin},\ and\ \citenamefont {Ihn}}]{basset_evaluating_2014}%
  \BibitemOpen
  \bibfield  {author} {\bibinfo {author} {\bibfnamefont {J.}~\bibnamefont {Basset}}, \bibinfo {author} {\bibfnamefont {A.}~\bibnamefont {Stockklauser}}, \bibinfo {author} {\bibfnamefont {D.-D.}\ \bibnamefont {Jarausch}}, \bibinfo {author} {\bibfnamefont {T.}~\bibnamefont {Frey}}, \bibinfo {author} {\bibfnamefont {C.}~\bibnamefont {Reichl}}, \bibinfo {author} {\bibfnamefont {W.}~\bibnamefont {Wegscheider}}, \bibinfo {author} {\bibfnamefont {A.}~\bibnamefont {Wallraff}}, \bibinfo {author} {\bibfnamefont {K.}~\bibnamefont {Ensslin}},\ and\ \bibinfo {author} {\bibfnamefont {T.}~\bibnamefont {Ihn}},\ }\bibfield  {title} {\bibinfo {title} {Evaluating charge noise acting on semiconductor quantum dots in the circuit quantum electrodynamics architecture},\ }\href {https://doi.org/10.1063/1.4892828} {\ \textbf {\bibinfo {volume} {105}},\ \bibinfo {pages} {063105}}\BibitemShut {NoStop}%
\bibitem [{\citenamefont {Dial}\ \emph {et~al.}()\citenamefont {Dial}, \citenamefont {Shulman}, \citenamefont {Harvey}, \citenamefont {Bluhm}, \citenamefont {Umansky},\ and\ \citenamefont {Yacoby}}]{dial_charge_2013}%
  \BibitemOpen
  \bibfield  {author} {\bibinfo {author} {\bibfnamefont {O.~E.}\ \bibnamefont {Dial}}, \bibinfo {author} {\bibfnamefont {M.~D.}\ \bibnamefont {Shulman}}, \bibinfo {author} {\bibfnamefont {S.~P.}\ \bibnamefont {Harvey}}, \bibinfo {author} {\bibfnamefont {H.}~\bibnamefont {Bluhm}}, \bibinfo {author} {\bibfnamefont {V.}~\bibnamefont {Umansky}},\ and\ \bibinfo {author} {\bibfnamefont {A.}~\bibnamefont {Yacoby}},\ }\bibfield  {title} {\bibinfo {title} {Charge noise spectroscopy using coherent exchange oscillations in a singlet-triplet qubit},\ }\href {https://doi.org/10.1103/PhysRevLett.110.146804} {\ \textbf {\bibinfo {volume} {110}},\ \bibinfo {pages} {146804}}\BibitemShut {NoStop}%
\bibitem [{\citenamefont {Jung}\ \emph {et~al.}()\citenamefont {Jung}, \citenamefont {Fujisawa}, \citenamefont {Hirayama},\ and\ \citenamefont {Jeong}}]{jung_background_2004}%
  \BibitemOpen
  \bibfield  {author} {\bibinfo {author} {\bibfnamefont {S.~W.}\ \bibnamefont {Jung}}, \bibinfo {author} {\bibfnamefont {T.}~\bibnamefont {Fujisawa}}, \bibinfo {author} {\bibfnamefont {Y.}~\bibnamefont {Hirayama}},\ and\ \bibinfo {author} {\bibfnamefont {Y.~H.}\ \bibnamefont {Jeong}},\ }\bibfield  {title} {\bibinfo {title} {Background charge fluctuation in a {GaAs} quantum dot device},\ }\href {https://doi.org/10.1063/1.1777802} {\ \textbf {\bibinfo {volume} {85}},\ \bibinfo {pages} {768}}\BibitemShut {NoStop}%
\bibitem [{\citenamefont {Elsayed}\ \emph {et~al.}()\citenamefont {Elsayed}, \citenamefont {Shehata}, \citenamefont {Godfrin}, \citenamefont {Kubicek}, \citenamefont {Massar}, \citenamefont {Canvel}, \citenamefont {Jussot}, \citenamefont {Simion}, \citenamefont {Mongillo}, \citenamefont {Wan}, \citenamefont {Govoreanu}, \citenamefont {Radu}, \citenamefont {Li}, \citenamefont {Van~Dorpe},\ and\ \citenamefont {De~Greve}}]{elsayed_low_2024}%
  \BibitemOpen
  \bibfield  {author} {\bibinfo {author} {\bibfnamefont {A.}~\bibnamefont {Elsayed}}, \bibinfo {author} {\bibfnamefont {M.~M.~K.}\ \bibnamefont {Shehata}}, \bibinfo {author} {\bibfnamefont {C.}~\bibnamefont {Godfrin}}, \bibinfo {author} {\bibfnamefont {S.}~\bibnamefont {Kubicek}}, \bibinfo {author} {\bibfnamefont {S.}~\bibnamefont {Massar}}, \bibinfo {author} {\bibfnamefont {Y.}~\bibnamefont {Canvel}}, \bibinfo {author} {\bibfnamefont {J.}~\bibnamefont {Jussot}}, \bibinfo {author} {\bibfnamefont {G.}~\bibnamefont {Simion}}, \bibinfo {author} {\bibfnamefont {M.}~\bibnamefont {Mongillo}}, \bibinfo {author} {\bibfnamefont {D.}~\bibnamefont {Wan}}, \bibinfo {author} {\bibfnamefont {B.}~\bibnamefont {Govoreanu}}, \bibinfo {author} {\bibfnamefont {I.~P.}\ \bibnamefont {Radu}}, \bibinfo {author} {\bibfnamefont {R.}~\bibnamefont {Li}}, \bibinfo {author} {\bibfnamefont {P.}~\bibnamefont {Van~Dorpe}},\ and\ \bibinfo {author} {\bibfnamefont {K.}~\bibnamefont {De~Greve}},\ }\bibfield  {title} {\bibinfo {title} {Low
  charge noise quantum dots with industrial {CMOS} manufacturing},\ }\href {https://doi.org/10.1038/s41534-024-00864-3} {\ \textbf {\bibinfo {volume} {10}},\ \bibinfo {pages} {70}}\BibitemShut {NoStop}%
\bibitem [{\citenamefont {Freeman}\ \emph {et~al.}()\citenamefont {Freeman}, \citenamefont {Schoenfield},\ and\ \citenamefont {Jiang}}]{freeman_comparison_2016}%
  \BibitemOpen
  \bibfield  {author} {\bibinfo {author} {\bibfnamefont {B.~M.}\ \bibnamefont {Freeman}}, \bibinfo {author} {\bibfnamefont {J.~S.}\ \bibnamefont {Schoenfield}},\ and\ \bibinfo {author} {\bibfnamefont {H.}~\bibnamefont {Jiang}},\ }\bibfield  {title} {\bibinfo {title} {Comparison of low frequency charge noise in identically patterned si/{SiO}2 and si/{SiGe} quantum dots},\ }\href {https://doi.org/10.1063/1.4954700} {\ \textbf {\bibinfo {volume} {108}},\ \bibinfo {pages} {253108}}\BibitemShut {NoStop}%
\bibitem [{\citenamefont {Jock}\ \emph {et~al.}()\citenamefont {Jock}, \citenamefont {Jacobson}, \citenamefont {Rudolph}, \citenamefont {Ward}, \citenamefont {Carroll},\ and\ \citenamefont {Luhman}}]{jock_silicon_2022}%
  \BibitemOpen
  \bibfield  {author} {\bibinfo {author} {\bibfnamefont {R.~M.}\ \bibnamefont {Jock}}, \bibinfo {author} {\bibfnamefont {N.~T.}\ \bibnamefont {Jacobson}}, \bibinfo {author} {\bibfnamefont {M.}~\bibnamefont {Rudolph}}, \bibinfo {author} {\bibfnamefont {D.~R.}\ \bibnamefont {Ward}}, \bibinfo {author} {\bibfnamefont {M.~S.}\ \bibnamefont {Carroll}},\ and\ \bibinfo {author} {\bibfnamefont {D.~R.}\ \bibnamefont {Luhman}},\ }\bibfield  {title} {\bibinfo {title} {A silicon singlet--triplet qubit driven by spin-valley coupling},\ }\href {https://doi.org/10.1038/s41467-022-28302-y} {\ \textbf {\bibinfo {volume} {13}},\ \bibinfo {pages} {641}}\BibitemShut {NoStop}%
\bibitem [{\citenamefont {Kim}\ \emph {et~al.}()\citenamefont {Kim}, \citenamefont {Hazard}, \citenamefont {Houck},\ and\ \citenamefont {Lyon}}]{kim_low-disorder_2019}%
  \BibitemOpen
  \bibfield  {author} {\bibinfo {author} {\bibfnamefont {J.-S.}\ \bibnamefont {Kim}}, \bibinfo {author} {\bibfnamefont {T.~M.}\ \bibnamefont {Hazard}}, \bibinfo {author} {\bibfnamefont {A.~A.}\ \bibnamefont {Houck}},\ and\ \bibinfo {author} {\bibfnamefont {S.~A.}\ \bibnamefont {Lyon}},\ }\bibfield  {title} {\bibinfo {title} {A low-disorder metal-oxide-silicon double quantum dot},\ }\href {https://doi.org/10.1063/1.5075486} {\ \textbf {\bibinfo {volume} {114}},\ \bibinfo {pages} {043501}}\BibitemShut {NoStop}%
\bibitem [{\citenamefont {Petit}\ \emph {et~al.}()\citenamefont {Petit}, \citenamefont {Boter}, \citenamefont {Eenink}, \citenamefont {Droulers}, \citenamefont {Tagliaferri}, \citenamefont {Li}, \citenamefont {Franke}, \citenamefont {Singh}, \citenamefont {Clarke}, \citenamefont {Schouten}, \citenamefont {Dobrovitski}, \citenamefont {Vandersypen},\ and\ \citenamefont {Veldhorst}}]{petit_spin_2018}%
  \BibitemOpen
  \bibfield  {author} {\bibinfo {author} {\bibfnamefont {L.}~\bibnamefont {Petit}}, \bibinfo {author} {\bibfnamefont {J.~M.}\ \bibnamefont {Boter}}, \bibinfo {author} {\bibfnamefont {H.~G.~J.}\ \bibnamefont {Eenink}}, \bibinfo {author} {\bibfnamefont {G.}~\bibnamefont {Droulers}}, \bibinfo {author} {\bibfnamefont {M.~L.~V.}\ \bibnamefont {Tagliaferri}}, \bibinfo {author} {\bibfnamefont {R.}~\bibnamefont {Li}}, \bibinfo {author} {\bibfnamefont {D.~P.}\ \bibnamefont {Franke}}, \bibinfo {author} {\bibfnamefont {K.~J.}\ \bibnamefont {Singh}}, \bibinfo {author} {\bibfnamefont {J.~S.}\ \bibnamefont {Clarke}}, \bibinfo {author} {\bibfnamefont {R.~N.}\ \bibnamefont {Schouten}}, \bibinfo {author} {\bibfnamefont {V.~V.}\ \bibnamefont {Dobrovitski}}, \bibinfo {author} {\bibfnamefont {L.~M.~K.}\ \bibnamefont {Vandersypen}},\ and\ \bibinfo {author} {\bibfnamefont {M.}~\bibnamefont {Veldhorst}},\ }\bibfield  {title} {\bibinfo {title} {Spin lifetime and charge noise in hot silicon quantum dot qubits},\ }\href
  {https://doi.org/10.1103/PhysRevLett.121.076801} {\ \textbf {\bibinfo {volume} {121}},\ \bibinfo {pages} {076801}}\BibitemShut {NoStop}%
\bibitem [{\citenamefont {Rudolph}\ \emph {et~al.}()\citenamefont {Rudolph}, \citenamefont {Sarabi}, \citenamefont {Murray}, \citenamefont {Carroll},\ and\ \citenamefont {Zimmerman}}]{rudolph_long-term_2019}%
  \BibitemOpen
  \bibfield  {author} {\bibinfo {author} {\bibfnamefont {M.}~\bibnamefont {Rudolph}}, \bibinfo {author} {\bibfnamefont {B.}~\bibnamefont {Sarabi}}, \bibinfo {author} {\bibfnamefont {R.}~\bibnamefont {Murray}}, \bibinfo {author} {\bibfnamefont {M.~S.}\ \bibnamefont {Carroll}},\ and\ \bibinfo {author} {\bibfnamefont {N.~M.}\ \bibnamefont {Zimmerman}},\ }\bibfield  {title} {\bibinfo {title} {Long-term drift of si-{MOS} quantum dots with intentional donor implants},\ }\href {https://doi.org/10.1038/s41598-019-43995-w} {\ \textbf {\bibinfo {volume} {9}},\ \bibinfo {pages} {7656}}\BibitemShut {NoStop}%
\bibitem [{\citenamefont {Spence}\ \emph {et~al.}()\citenamefont {Spence}, \citenamefont {Cardoso~Paz}, \citenamefont {Michal}, \citenamefont {Chanrion}, \citenamefont {Niegemann}, \citenamefont {Jadot}, \citenamefont {Mortemousque}, \citenamefont {Klemt}, \citenamefont {Thiney}, \citenamefont {Bertrand}, \citenamefont {Hutin}, \citenamefont {B{\"a}uerle}, \citenamefont {Vinet}, \citenamefont {Niquet}, \citenamefont {Meunier},\ and\ \citenamefont {Urdampilleta}}]{spence_probing_2023}%
  \BibitemOpen
  \bibfield  {author} {\bibinfo {author} {\bibfnamefont {C.}~\bibnamefont {Spence}}, \bibinfo {author} {\bibfnamefont {B.}~\bibnamefont {Cardoso~Paz}}, \bibinfo {author} {\bibfnamefont {V.}~\bibnamefont {Michal}}, \bibinfo {author} {\bibfnamefont {E.}~\bibnamefont {Chanrion}}, \bibinfo {author} {\bibfnamefont {D.~J.}\ \bibnamefont {Niegemann}}, \bibinfo {author} {\bibfnamefont {B.}~\bibnamefont {Jadot}}, \bibinfo {author} {\bibfnamefont {P.-A.}\ \bibnamefont {Mortemousque}}, \bibinfo {author} {\bibfnamefont {B.}~\bibnamefont {Klemt}}, \bibinfo {author} {\bibfnamefont {V.}~\bibnamefont {Thiney}}, \bibinfo {author} {\bibfnamefont {B.}~\bibnamefont {Bertrand}}, \bibinfo {author} {\bibfnamefont {L.}~\bibnamefont {Hutin}}, \bibinfo {author} {\bibfnamefont {C.}~\bibnamefont {B{\"a}uerle}}, \bibinfo {author} {\bibfnamefont {M.}~\bibnamefont {Vinet}}, \bibinfo {author} {\bibfnamefont {Y.-M.}\ \bibnamefont {Niquet}}, \bibinfo {author} {\bibfnamefont {T.}~\bibnamefont {Meunier}},\ and\ \bibinfo {author} {\bibfnamefont
  {M.}~\bibnamefont {Urdampilleta}},\ }\bibfield  {title} {\bibinfo {title} {Probing low-frequency charge noise in few-electron {CMOS} quantum dots},\ }\href {https://doi.org/10.1103/PhysRevApplied.19.044010} {\ \textbf {\bibinfo {volume} {19}},\ \bibinfo {pages} {044010}}\BibitemShut {NoStop}%
\bibitem [{\citenamefont {Stuyck}\ \emph {et~al.}()\citenamefont {Stuyck}, \citenamefont {Li}, \citenamefont {Godfrin}, \citenamefont {Elsayed}, \citenamefont {Kubicek}, \citenamefont {Jussot}, \citenamefont {Chan}, \citenamefont {Mohiyaddin}, \citenamefont {Shehata}, \citenamefont {Simion}, \citenamefont {Canvel}, \citenamefont {Goux}, \citenamefont {Heyns}, \citenamefont {Govoreanu},\ and\ \citenamefont {Radu}}]{stuyck_uniform_2021}%
  \BibitemOpen
  \bibfield  {author} {\bibinfo {author} {\bibfnamefont {N.~I.~D.}\ \bibnamefont {Stuyck}}, \bibinfo {author} {\bibfnamefont {R.}~\bibnamefont {Li}}, \bibinfo {author} {\bibfnamefont {C.}~\bibnamefont {Godfrin}}, \bibinfo {author} {\bibfnamefont {A.}~\bibnamefont {Elsayed}}, \bibinfo {author} {\bibfnamefont {S.}~\bibnamefont {Kubicek}}, \bibinfo {author} {\bibfnamefont {J.}~\bibnamefont {Jussot}}, \bibinfo {author} {\bibfnamefont {B.~T.}\ \bibnamefont {Chan}}, \bibinfo {author} {\bibfnamefont {F.~A.}\ \bibnamefont {Mohiyaddin}}, \bibinfo {author} {\bibfnamefont {M.}~\bibnamefont {Shehata}}, \bibinfo {author} {\bibfnamefont {G.}~\bibnamefont {Simion}}, \bibinfo {author} {\bibfnamefont {Y.}~\bibnamefont {Canvel}}, \bibinfo {author} {\bibfnamefont {L.}~\bibnamefont {Goux}}, \bibinfo {author} {\bibfnamefont {M.}~\bibnamefont {Heyns}}, \bibinfo {author} {\bibfnamefont {B.}~\bibnamefont {Govoreanu}},\ and\ \bibinfo {author} {\bibfnamefont {I.~P.}\ \bibnamefont {Radu}},\ }\bibfield  {title} {\bibinfo {title}
  {Uniform spin qubit devices with tunable coupling in an all-silicon 300 mm integrated process},\ }in\ \href {https://doi.org/10.23919/VLSICircuits52068.2021.9492427} {\emph {\bibinfo {booktitle} {2021 Symposium on {VLSI} Circuits}}},\ pp.\ \bibinfo {pages} {1--2},\ \bibinfo {note} {{ISSN}: 2158-5636}\BibitemShut {NoStop}%
\bibitem [{\citenamefont {Tomi{\'c}}\ \emph {et~al.}()\citenamefont {Tomi{\'c}}, \citenamefont {B{\"u}tler}, \citenamefont {Wu}, \citenamefont {Raes}, \citenamefont {Godfrin}, \citenamefont {Kubicek}, \citenamefont {Jussot}, \citenamefont {Canvel}, \citenamefont {Hermans}, \citenamefont {Shimura}, \citenamefont {Loo}, \citenamefont {Beyne}, \citenamefont {Jaliel}, \citenamefont {Caekenberghe}, \citenamefont {Levajac}, \citenamefont {Wan}, \citenamefont {Greve}, \citenamefont {Huang}, \citenamefont {Ensslin},\ and\ \citenamefont {Ihn}}]{tomic_long_2025}%
  \BibitemOpen
  \bibfield  {author} {\bibinfo {author} {\bibfnamefont {P.}~\bibnamefont {Tomi{\'c}}}, \bibinfo {author} {\bibfnamefont {P.}~\bibnamefont {B{\"u}tler}}, \bibinfo {author} {\bibfnamefont {Y.}~\bibnamefont {Wu}}, \bibinfo {author} {\bibfnamefont {B.}~\bibnamefont {Raes}}, \bibinfo {author} {\bibfnamefont {C.}~\bibnamefont {Godfrin}}, \bibinfo {author} {\bibfnamefont {S.}~\bibnamefont {Kubicek}}, \bibinfo {author} {\bibfnamefont {J.}~\bibnamefont {Jussot}}, \bibinfo {author} {\bibfnamefont {Y.}~\bibnamefont {Canvel}}, \bibinfo {author} {\bibfnamefont {Y.}~\bibnamefont {Hermans}}, \bibinfo {author} {\bibfnamefont {Y.}~\bibnamefont {Shimura}}, \bibinfo {author} {\bibfnamefont {R.}~\bibnamefont {Loo}}, \bibinfo {author} {\bibfnamefont {S.}~\bibnamefont {Beyne}}, \bibinfo {author} {\bibfnamefont {G.}~\bibnamefont {Jaliel}}, \bibinfo {author} {\bibfnamefont {T.~V.}\ \bibnamefont {Caekenberghe}}, \bibinfo {author} {\bibfnamefont {V.}~\bibnamefont {Levajac}}, \bibinfo {author} {\bibfnamefont {D.}~\bibnamefont {Wan}},
  \bibinfo {author} {\bibfnamefont {K.~D.}\ \bibnamefont {Greve}}, \bibinfo {author} {\bibfnamefont {W.~W.}\ \bibnamefont {Huang}}, \bibinfo {author} {\bibfnamefont {K.}~\bibnamefont {Ensslin}},\ and\ \bibinfo {author} {\bibfnamefont {T.}~\bibnamefont {Ihn}},\ }\href {https://doi.org/10.48550/arXiv.2512.20758} {\bibinfo {title} {Long coherence silicon spin qubit fabricated in a 300 mm industrial foundry}},\ \Eprint {https://arxiv.org/abs/2512.20758 [cond-mat.mes-hall]} {2512.20758 [cond-mat.mes-hall]} \BibitemShut {NoStop}%
\bibitem [{\citenamefont {Zwerver}\ \emph {et~al.}()\citenamefont {Zwerver}, \citenamefont {Kr{\"a}henmann}, \citenamefont {Watson}, \citenamefont {Lampert}, \citenamefont {George}, \citenamefont {Pillarisetty}, \citenamefont {Bojarski}, \citenamefont {Amin}, \citenamefont {Amitonov}, \citenamefont {Boter}, \citenamefont {Caudillo}, \citenamefont {Correas-Serrano}, \citenamefont {Dehollain}, \citenamefont {Droulers}, \citenamefont {Henry}, \citenamefont {Kotlyar}, \citenamefont {Lodari}, \citenamefont {L{\"u}thi}, \citenamefont {Michalak}, \citenamefont {Mueller}, \citenamefont {Neyens}, \citenamefont {Roberts}, \citenamefont {Samkharadze}, \citenamefont {Zheng}, \citenamefont {Zietz}, \citenamefont {Scappucci}, \citenamefont {Veldhorst}, \citenamefont {Vandersypen},\ and\ \citenamefont {Clarke}}]{zwerver_qubits_2022}%
  \BibitemOpen
  \bibfield  {author} {\bibinfo {author} {\bibfnamefont {A.~M.~J.}\ \bibnamefont {Zwerver}}, \bibinfo {author} {\bibfnamefont {T.}~\bibnamefont {Kr{\"a}henmann}}, \bibinfo {author} {\bibfnamefont {T.~F.}\ \bibnamefont {Watson}}, \bibinfo {author} {\bibfnamefont {L.}~\bibnamefont {Lampert}}, \bibinfo {author} {\bibfnamefont {H.~C.}\ \bibnamefont {George}}, \bibinfo {author} {\bibfnamefont {R.}~\bibnamefont {Pillarisetty}}, \bibinfo {author} {\bibfnamefont {S.~A.}\ \bibnamefont {Bojarski}}, \bibinfo {author} {\bibfnamefont {P.}~\bibnamefont {Amin}}, \bibinfo {author} {\bibfnamefont {S.~V.}\ \bibnamefont {Amitonov}}, \bibinfo {author} {\bibfnamefont {J.~M.}\ \bibnamefont {Boter}}, \bibinfo {author} {\bibfnamefont {R.}~\bibnamefont {Caudillo}}, \bibinfo {author} {\bibfnamefont {D.}~\bibnamefont {Correas-Serrano}}, \bibinfo {author} {\bibfnamefont {J.~P.}\ \bibnamefont {Dehollain}}, \bibinfo {author} {\bibfnamefont {G.}~\bibnamefont {Droulers}}, \bibinfo {author} {\bibfnamefont {E.~M.}\ \bibnamefont {Henry}},
  \bibinfo {author} {\bibfnamefont {R.}~\bibnamefont {Kotlyar}}, \bibinfo {author} {\bibfnamefont {M.}~\bibnamefont {Lodari}}, \bibinfo {author} {\bibfnamefont {F.}~\bibnamefont {L{\"u}thi}}, \bibinfo {author} {\bibfnamefont {D.~J.}\ \bibnamefont {Michalak}}, \bibinfo {author} {\bibfnamefont {B.~K.}\ \bibnamefont {Mueller}}, \bibinfo {author} {\bibfnamefont {S.}~\bibnamefont {Neyens}}, \bibinfo {author} {\bibfnamefont {J.}~\bibnamefont {Roberts}}, \bibinfo {author} {\bibfnamefont {N.}~\bibnamefont {Samkharadze}}, \bibinfo {author} {\bibfnamefont {G.}~\bibnamefont {Zheng}}, \bibinfo {author} {\bibfnamefont {O.~K.}\ \bibnamefont {Zietz}}, \bibinfo {author} {\bibfnamefont {G.}~\bibnamefont {Scappucci}}, \bibinfo {author} {\bibfnamefont {M.}~\bibnamefont {Veldhorst}}, \bibinfo {author} {\bibfnamefont {L.~M.~K.}\ \bibnamefont {Vandersypen}},\ and\ \bibinfo {author} {\bibfnamefont {J.~S.}\ \bibnamefont {Clarke}},\ }\bibfield  {title} {\bibinfo {title} {Qubits made by advanced semiconductor manufacturing},\ }\href
  {https://doi.org/10.1038/s41928-022-00727-9} {\ \textbf {\bibinfo {volume} {5}},\ \bibinfo {pages} {184}}\BibitemShut {NoStop}%
\bibitem [{\citenamefont {Connors}\ \emph {et~al.}({\natexlab{b}})\citenamefont {Connors}, \citenamefont {Nelson}, \citenamefont {Edge},\ and\ \citenamefont {Nichol}}]{connors_charge-noise_2022}%
  \BibitemOpen
  \bibfield  {author} {\bibinfo {author} {\bibfnamefont {E.~J.}\ \bibnamefont {Connors}}, \bibinfo {author} {\bibfnamefont {J.}~\bibnamefont {Nelson}}, \bibinfo {author} {\bibfnamefont {L.~F.}\ \bibnamefont {Edge}},\ and\ \bibinfo {author} {\bibfnamefont {J.~M.}\ \bibnamefont {Nichol}},\ }\bibfield  {title} {\bibinfo {title} {Charge-noise spectroscopy of si/{SiGe} quantum dots via dynamically-decoupled exchange oscillations},\ }\href {https://doi.org/10.1038/s41467-022-28519-x} {\ \textbf {\bibinfo {volume} {13}},\ \bibinfo {pages} {940} ({\natexlab{b}})}\BibitemShut {NoStop}%
\bibitem [{\citenamefont {Degli~Esposti}\ \emph {et~al.}()\citenamefont {Degli~Esposti}, \citenamefont {Stehouwer}, \citenamefont {G{\"u}l}, \citenamefont {Samkharadze}, \citenamefont {D{\'e}prez}, \citenamefont {Meyer}, \citenamefont {Meijer}, \citenamefont {Tryputen}, \citenamefont {Karwal}, \citenamefont {Botifoll}, \citenamefont {Arbiol}, \citenamefont {Amitonov}, \citenamefont {Vandersypen}, \citenamefont {Sammak}, \citenamefont {Veldhorst},\ and\ \citenamefont {Scappucci}}]{degli_esposti_low_2024}%
  \BibitemOpen
  \bibfield  {author} {\bibinfo {author} {\bibfnamefont {D.}~\bibnamefont {Degli~Esposti}}, \bibinfo {author} {\bibfnamefont {L.~E.~A.}\ \bibnamefont {Stehouwer}}, \bibinfo {author} {\bibfnamefont {{\"O}.}~\bibnamefont {G{\"u}l}}, \bibinfo {author} {\bibfnamefont {N.}~\bibnamefont {Samkharadze}}, \bibinfo {author} {\bibfnamefont {C.}~\bibnamefont {D{\'e}prez}}, \bibinfo {author} {\bibfnamefont {M.}~\bibnamefont {Meyer}}, \bibinfo {author} {\bibfnamefont {I.~N.}\ \bibnamefont {Meijer}}, \bibinfo {author} {\bibfnamefont {L.}~\bibnamefont {Tryputen}}, \bibinfo {author} {\bibfnamefont {S.}~\bibnamefont {Karwal}}, \bibinfo {author} {\bibfnamefont {M.}~\bibnamefont {Botifoll}}, \bibinfo {author} {\bibfnamefont {J.}~\bibnamefont {Arbiol}}, \bibinfo {author} {\bibfnamefont {S.~V.}\ \bibnamefont {Amitonov}}, \bibinfo {author} {\bibfnamefont {L.~M.~K.}\ \bibnamefont {Vandersypen}}, \bibinfo {author} {\bibfnamefont {A.}~\bibnamefont {Sammak}}, \bibinfo {author} {\bibfnamefont {M.}~\bibnamefont {Veldhorst}},\ and\
  \bibinfo {author} {\bibfnamefont {G.}~\bibnamefont {Scappucci}},\ }\bibfield  {title} {\bibinfo {title} {Low disorder and high valley splitting in silicon},\ }\href {https://doi.org/10.1038/s41534-024-00826-9} {\ \textbf {\bibinfo {volume} {10}},\ \bibinfo {pages} {32}}\BibitemShut {NoStop}%
\bibitem [{\citenamefont {Mi}\ \emph {et~al.}()\citenamefont {Mi}, \citenamefont {Kohler},\ and\ \citenamefont {Petta}}]{mi_landau-zener_2018}%
  \BibitemOpen
  \bibfield  {author} {\bibinfo {author} {\bibfnamefont {X.}~\bibnamefont {Mi}}, \bibinfo {author} {\bibfnamefont {S.}~\bibnamefont {Kohler}},\ and\ \bibinfo {author} {\bibfnamefont {J.~R.}\ \bibnamefont {Petta}},\ }\bibfield  {title} {\bibinfo {title} {Landau-zener interferometry of valley-orbit states in si/{SiGe} double quantum dots},\ }\href {https://doi.org/10.1103/PhysRevB.98.161404} {\ \textbf {\bibinfo {volume} {98}},\ \bibinfo {pages} {161404}}\BibitemShut {NoStop}%
\bibitem [{\citenamefont {Paquelet~Wuetz}\ \emph {et~al.}()\citenamefont {Paquelet~Wuetz}, \citenamefont {Degli~Esposti}, \citenamefont {Zwerver}, \citenamefont {Amitonov}, \citenamefont {Botifoll}, \citenamefont {Arbiol}, \citenamefont {Sammak}, \citenamefont {Vandersypen}, \citenamefont {Russ},\ and\ \citenamefont {Scappucci}}]{paquelet_wuetz_reducing_2023}%
  \BibitemOpen
  \bibfield  {author} {\bibinfo {author} {\bibfnamefont {B.}~\bibnamefont {Paquelet~Wuetz}}, \bibinfo {author} {\bibfnamefont {D.}~\bibnamefont {Degli~Esposti}}, \bibinfo {author} {\bibfnamefont {A.-M.~J.}\ \bibnamefont {Zwerver}}, \bibinfo {author} {\bibfnamefont {S.~V.}\ \bibnamefont {Amitonov}}, \bibinfo {author} {\bibfnamefont {M.}~\bibnamefont {Botifoll}}, \bibinfo {author} {\bibfnamefont {J.}~\bibnamefont {Arbiol}}, \bibinfo {author} {\bibfnamefont {A.}~\bibnamefont {Sammak}}, \bibinfo {author} {\bibfnamefont {L.~M.~K.}\ \bibnamefont {Vandersypen}}, \bibinfo {author} {\bibfnamefont {M.}~\bibnamefont {Russ}},\ and\ \bibinfo {author} {\bibfnamefont {G.}~\bibnamefont {Scappucci}},\ }\bibfield  {title} {\bibinfo {title} {Reducing charge noise in quantum dots by using thin silicon quantum wells},\ }\href {https://doi.org/10.1038/s41467-023-36951-w} {\ \textbf {\bibinfo {volume} {14}},\ \bibinfo {pages} {1385}}\BibitemShut {NoStop}%
\bibitem [{\citenamefont {Struck}\ \emph {et~al.}()\citenamefont {Struck}, \citenamefont {Hollmann}, \citenamefont {Schauer}, \citenamefont {Fedorets}, \citenamefont {Schmidbauer}, \citenamefont {Sawano}, \citenamefont {Riemann}, \citenamefont {Abrosimov}, \citenamefont {Cywi{\'n}ski}, \citenamefont {Bougeard},\ and\ \citenamefont {Schreiber}}]{struck_low-frequency_2020}%
  \BibitemOpen
  \bibfield  {author} {\bibinfo {author} {\bibfnamefont {T.}~\bibnamefont {Struck}}, \bibinfo {author} {\bibfnamefont {A.}~\bibnamefont {Hollmann}}, \bibinfo {author} {\bibfnamefont {F.}~\bibnamefont {Schauer}}, \bibinfo {author} {\bibfnamefont {O.}~\bibnamefont {Fedorets}}, \bibinfo {author} {\bibfnamefont {A.}~\bibnamefont {Schmidbauer}}, \bibinfo {author} {\bibfnamefont {K.}~\bibnamefont {Sawano}}, \bibinfo {author} {\bibfnamefont {H.}~\bibnamefont {Riemann}}, \bibinfo {author} {\bibfnamefont {N.~V.}\ \bibnamefont {Abrosimov}}, \bibinfo {author} {\bibfnamefont {{\L}.}~\bibnamefont {Cywi{\'n}ski}}, \bibinfo {author} {\bibfnamefont {D.}~\bibnamefont {Bougeard}},\ and\ \bibinfo {author} {\bibfnamefont {L.~R.}\ \bibnamefont {Schreiber}},\ }\bibfield  {title} {\bibinfo {title} {Low-frequency spin qubit energy splitting noise in highly purified 28si/{SiGe}},\ }\href {https://doi.org/10.1038/s41534-020-0276-2} {\ \textbf {\bibinfo {volume} {6}},\ \bibinfo {pages} {40}}\BibitemShut {NoStop}%
\bibitem [{\citenamefont {Hendrickx}\ \emph {et~al.}()\citenamefont {Hendrickx}, \citenamefont {Massai}, \citenamefont {Mergenthaler}, \citenamefont {Schupp}, \citenamefont {Paredes}, \citenamefont {Bedell}, \citenamefont {Salis},\ and\ \citenamefont {Fuhrer}}]{hendrickx_sweet-spot_2024}%
  \BibitemOpen
  \bibfield  {author} {\bibinfo {author} {\bibfnamefont {N.~W.}\ \bibnamefont {Hendrickx}}, \bibinfo {author} {\bibfnamefont {L.}~\bibnamefont {Massai}}, \bibinfo {author} {\bibfnamefont {M.}~\bibnamefont {Mergenthaler}}, \bibinfo {author} {\bibfnamefont {F.~J.}\ \bibnamefont {Schupp}}, \bibinfo {author} {\bibfnamefont {S.}~\bibnamefont {Paredes}}, \bibinfo {author} {\bibfnamefont {S.~W.}\ \bibnamefont {Bedell}}, \bibinfo {author} {\bibfnamefont {G.}~\bibnamefont {Salis}},\ and\ \bibinfo {author} {\bibfnamefont {A.}~\bibnamefont {Fuhrer}},\ }\bibfield  {title} {\bibinfo {title} {Sweet-spot operation of a germanium hole spin qubit with highly anisotropic noise sensitivity},\ }\href {https://doi.org/10.1038/s41563-024-01857-5} {\ \textbf {\bibinfo {volume} {23}},\ \bibinfo {pages} {920}}\BibitemShut {NoStop}%
\bibitem [{\citenamefont {Lodari}\ \emph {et~al.}()\citenamefont {Lodari}, \citenamefont {Hendrickx}, \citenamefont {Lawrie}, \citenamefont {Hsiao}, \citenamefont {Vandersypen}, \citenamefont {Sammak}, \citenamefont {Veldhorst},\ and\ \citenamefont {Scappucci}}]{lodari_low_2021}%
  \BibitemOpen
  \bibfield  {author} {\bibinfo {author} {\bibfnamefont {M.}~\bibnamefont {Lodari}}, \bibinfo {author} {\bibfnamefont {N.~W.}\ \bibnamefont {Hendrickx}}, \bibinfo {author} {\bibfnamefont {W.~I.~L.}\ \bibnamefont {Lawrie}}, \bibinfo {author} {\bibfnamefont {T.-K.}\ \bibnamefont {Hsiao}}, \bibinfo {author} {\bibfnamefont {L.~M.~K.}\ \bibnamefont {Vandersypen}}, \bibinfo {author} {\bibfnamefont {A.}~\bibnamefont {Sammak}}, \bibinfo {author} {\bibfnamefont {M.}~\bibnamefont {Veldhorst}},\ and\ \bibinfo {author} {\bibfnamefont {G.}~\bibnamefont {Scappucci}},\ }\bibfield  {title} {\bibinfo {title} {Low percolation density and charge noise with holes in germanium},\ }\href {https://doi.org/10.1088/2633-4356/abcd82} {\ \textbf {\bibinfo {volume} {1}},\ \bibinfo {pages} {011002}}\BibitemShut {NoStop}%
\bibitem [{\citenamefont {Yoneda}\ \emph {et~al.}()\citenamefont {Yoneda}, \citenamefont {Rojas-Arias}, \citenamefont {Stano}, \citenamefont {Takeda}, \citenamefont {Noiri}, \citenamefont {Nakajima}, \citenamefont {Loss},\ and\ \citenamefont {Tarucha}}]{yoneda_noise-correlation_2023}%
  \BibitemOpen
  \bibfield  {author} {\bibinfo {author} {\bibfnamefont {J.}~\bibnamefont {Yoneda}}, \bibinfo {author} {\bibfnamefont {J.~S.}\ \bibnamefont {Rojas-Arias}}, \bibinfo {author} {\bibfnamefont {P.}~\bibnamefont {Stano}}, \bibinfo {author} {\bibfnamefont {K.}~\bibnamefont {Takeda}}, \bibinfo {author} {\bibfnamefont {A.}~\bibnamefont {Noiri}}, \bibinfo {author} {\bibfnamefont {T.}~\bibnamefont {Nakajima}}, \bibinfo {author} {\bibfnamefont {D.}~\bibnamefont {Loss}},\ and\ \bibinfo {author} {\bibfnamefont {S.}~\bibnamefont {Tarucha}},\ }\bibfield  {title} {\bibinfo {title} {Noise-correlation spectrum for a pair of spin qubits in silicon},\ }\href {https://doi.org/10.1038/s41567-023-02238-6} {\ \textbf {\bibinfo {volume} {19}},\ \bibinfo {pages} {1793}}\BibitemShut {NoStop}%
\bibitem [{\citenamefont {Ruckriegel}\ \emph {et~al.}()\citenamefont {Ruckriegel}, \citenamefont {Adam}, \citenamefont {Bolt}, \citenamefont {Tong}, \citenamefont {Kealhofer}, \citenamefont {Denisov}, \citenamefont {Panah}, \citenamefont {Watanabe}, \citenamefont {Taniguchi}, \citenamefont {Ihn},\ and\ \citenamefont {Ensslin}}]{ruckriegel_microwave_2026}%
  \BibitemOpen
  \bibfield  {author} {\bibinfo {author} {\bibfnamefont {M.~J.}\ \bibnamefont {Ruckriegel}}, \bibinfo {author} {\bibfnamefont {C.}~\bibnamefont {Adam}}, \bibinfo {author} {\bibfnamefont {R.}~\bibnamefont {Bolt}}, \bibinfo {author} {\bibfnamefont {C.}~\bibnamefont {Tong}}, \bibinfo {author} {\bibfnamefont {D.}~\bibnamefont {Kealhofer}}, \bibinfo {author} {\bibfnamefont {A.~O.}\ \bibnamefont {Denisov}}, \bibinfo {author} {\bibfnamefont {M.~B.}\ \bibnamefont {Panah}}, \bibinfo {author} {\bibfnamefont {K.}~\bibnamefont {Watanabe}}, \bibinfo {author} {\bibfnamefont {T.}~\bibnamefont {Taniguchi}}, \bibinfo {author} {\bibfnamefont {T.}~\bibnamefont {Ihn}},\ and\ \bibinfo {author} {\bibfnamefont {K.}~\bibnamefont {Ensslin}},\ }\bibfield  {title} {\bibinfo {title} {Microwave spectroscopy of few-carrier states in bilayer graphene quantum dots},\ }\href {https://doi.org/10.1103/j1ts-9nys} {\ \textbf {\bibinfo {volume} {7}},\ \bibinfo {pages} {033037}}\BibitemShut {NoStop}%
\bibitem [{\citenamefont {Rumyantsev}\ \emph {et~al.}()\citenamefont {Rumyantsev}, \citenamefont {Jiang}, \citenamefont {Samnakay}, \citenamefont {Shur},\ and\ \citenamefont {Balandin}}]{rumyantsev_1_2015}%
  \BibitemOpen
  \bibfield  {author} {\bibinfo {author} {\bibfnamefont {S.~L.}\ \bibnamefont {Rumyantsev}}, \bibinfo {author} {\bibfnamefont {C.}~\bibnamefont {Jiang}}, \bibinfo {author} {\bibfnamefont {R.}~\bibnamefont {Samnakay}}, \bibinfo {author} {\bibfnamefont {M.~S.}\ \bibnamefont {Shur}},\ and\ \bibinfo {author} {\bibfnamefont {A.~A.}\ \bibnamefont {Balandin}},\ }\bibfield  {title} {\bibinfo {title} {1/ f noise characteristics of {MoS}2 thin-film transistors: Comparison of single and multilayer structures},\ }\href {https://doi.org/10.1109/LED.2015.2412536} {\ \textbf {\bibinfo {volume} {36}},\ \bibinfo {pages} {517}}\BibitemShut {NoStop}%
\bibitem [{\citenamefont {Ko}\ \emph {et~al.}()\citenamefont {Ko}, \citenamefont {Shin}, \citenamefont {Kim}, \citenamefont {Jang}, \citenamefont {Jin}, \citenamefont {Shin}, \citenamefont {Kim},\ and\ \citenamefont {Kim}}]{ko_current_2015}%
  \BibitemOpen
  \bibfield  {author} {\bibinfo {author} {\bibfnamefont {S.-P.}\ \bibnamefont {Ko}}, \bibinfo {author} {\bibfnamefont {J.~M.}\ \bibnamefont {Shin}}, \bibinfo {author} {\bibfnamefont {Y.~J.}\ \bibnamefont {Kim}}, \bibinfo {author} {\bibfnamefont {H.-K.}\ \bibnamefont {Jang}}, \bibinfo {author} {\bibfnamefont {J.~E.}\ \bibnamefont {Jin}}, \bibinfo {author} {\bibfnamefont {M.}~\bibnamefont {Shin}}, \bibinfo {author} {\bibfnamefont {Y.~K.}\ \bibnamefont {Kim}},\ and\ \bibinfo {author} {\bibfnamefont {G.-T.}\ \bibnamefont {Kim}},\ }\bibfield  {title} {\bibinfo {title} {Current fluctuation of electron and hole carriers in multilayer {WSe}2 field effect transistors},\ }\href {https://doi.org/10.1063/1.4937618} {\ \textbf {\bibinfo {volume} {107}},\ \bibinfo {pages} {242102}}\BibitemShut {NoStop}%
\bibitem [{\citenamefont {Gerber}\ \emph {et~al.}({\natexlab{b}})\citenamefont {Gerber}, \citenamefont {Ersoy}, \citenamefont {Masseroni}, \citenamefont {Niese}, \citenamefont {Denisov}, \citenamefont {Adam}, \citenamefont {Ostertag}, \citenamefont {Richter}, \citenamefont {Taniguchi}, \citenamefont {Watanabe}, \citenamefont {Meir}, \citenamefont {Ihn},\ and\ \citenamefont {Ensslin}}]{gerber_spin-valley_2026}%
  \BibitemOpen
  \bibfield  {author} {\bibinfo {author} {\bibfnamefont {J.~D.}\ \bibnamefont {Gerber}}, \bibinfo {author} {\bibfnamefont {E.}~\bibnamefont {Ersoy}}, \bibinfo {author} {\bibfnamefont {M.}~\bibnamefont {Masseroni}}, \bibinfo {author} {\bibfnamefont {M.}~\bibnamefont {Niese}}, \bibinfo {author} {\bibfnamefont {A.~O.}\ \bibnamefont {Denisov}}, \bibinfo {author} {\bibfnamefont {C.}~\bibnamefont {Adam}}, \bibinfo {author} {\bibfnamefont {L.}~\bibnamefont {Ostertag}}, \bibinfo {author} {\bibfnamefont {J.}~\bibnamefont {Richter}}, \bibinfo {author} {\bibfnamefont {T.}~\bibnamefont {Taniguchi}}, \bibinfo {author} {\bibfnamefont {K.}~\bibnamefont {Watanabe}}, \bibinfo {author} {\bibfnamefont {Y.}~\bibnamefont {Meir}}, \bibinfo {author} {\bibfnamefont {T.}~\bibnamefont {Ihn}},\ and\ \bibinfo {author} {\bibfnamefont {K.}~\bibnamefont {Ensslin}},\ }\bibfield  {title} {\bibinfo {title} {Spin-valley 0.7 anomaly in bilayer graphene/{WSe}2 quantum point contacts},\ }\href {https://doi.org/10.1103/zss4-lmlm} {\ \textbf
  {\bibinfo {volume} {114}},\ \bibinfo {pages} {L111401} ({\natexlab{b}})}\BibitemShut {NoStop}%
\bibitem [{\citenamefont {Shnirman}\ \emph {et~al.}()\citenamefont {Shnirman}, \citenamefont {Makhlin},\ and\ \citenamefont {Sch{\"o}n}}]{shnirman_noise_2002}%
  \BibitemOpen
  \bibfield  {author} {\bibinfo {author} {\bibfnamefont {A.}~\bibnamefont {Shnirman}}, \bibinfo {author} {\bibfnamefont {Y.}~\bibnamefont {Makhlin}},\ and\ \bibinfo {author} {\bibfnamefont {G.}~\bibnamefont {Sch{\"o}n}},\ }\bibfield  {title} {\bibinfo {title} {Noise and decoherence in quantum two-level systems},\ }\href {https://doi.org/10.1238/Physica.Topical.102a00147} {\ \textbf {\bibinfo {volume} {2002}},\ \bibinfo {pages} {147}}\BibitemShut {NoStop}%
\bibitem [{\citenamefont {Bermeister}\ \emph {et~al.}()\citenamefont {Bermeister}, \citenamefont {Keith},\ and\ \citenamefont {Culcer}}]{bermeister_charge_2014}%
  \BibitemOpen
  \bibfield  {author} {\bibinfo {author} {\bibfnamefont {A.}~\bibnamefont {Bermeister}}, \bibinfo {author} {\bibfnamefont {D.}~\bibnamefont {Keith}},\ and\ \bibinfo {author} {\bibfnamefont {D.}~\bibnamefont {Culcer}},\ }\bibfield  {title} {\bibinfo {title} {Charge noise, spin-orbit coupling, and dephasing of single-spin qubits},\ }\href {https://doi.org/10.1063/1.4901162} {\ \textbf {\bibinfo {volume} {105}},\ \bibinfo {pages} {192102}}\BibitemShut {NoStop}%
\bibitem [{\citenamefont {{MacQuarrie}}\ \emph {et~al.}()\citenamefont {{MacQuarrie}}, \citenamefont {Neyens}, \citenamefont {Dodson}, \citenamefont {Corrigan}, \citenamefont {Thorgrimsson}, \citenamefont {Holman}, \citenamefont {Palma}, \citenamefont {Edge}, \citenamefont {Friesen}, \citenamefont {Coppersmith},\ and\ \citenamefont {Eriksson}}]{macquarrie_progress_2020}%
  \BibitemOpen
  \bibfield  {author} {\bibinfo {author} {\bibfnamefont {E.~R.}\ \bibnamefont {{MacQuarrie}}}, \bibinfo {author} {\bibfnamefont {S.~F.}\ \bibnamefont {Neyens}}, \bibinfo {author} {\bibfnamefont {J.~P.}\ \bibnamefont {Dodson}}, \bibinfo {author} {\bibfnamefont {J.}~\bibnamefont {Corrigan}}, \bibinfo {author} {\bibfnamefont {B.}~\bibnamefont {Thorgrimsson}}, \bibinfo {author} {\bibfnamefont {N.}~\bibnamefont {Holman}}, \bibinfo {author} {\bibfnamefont {M.}~\bibnamefont {Palma}}, \bibinfo {author} {\bibfnamefont {L.~F.}\ \bibnamefont {Edge}}, \bibinfo {author} {\bibfnamefont {M.}~\bibnamefont {Friesen}}, \bibinfo {author} {\bibfnamefont {S.~N.}\ \bibnamefont {Coppersmith}},\ and\ \bibinfo {author} {\bibfnamefont {M.~A.}\ \bibnamefont {Eriksson}},\ }\bibfield  {title} {\bibinfo {title} {Progress toward a capacitively mediated {CNOT} between two charge qubits in si/{SiGe}},\ }\href {https://doi.org/10.1038/s41534-020-00314-w} {\ \textbf {\bibinfo {volume} {6}},\ \bibinfo {pages} {81}}\BibitemShut {NoStop}%
\bibitem [{\citenamefont {Viennot}\ \emph {et~al.}()\citenamefont {Viennot}, \citenamefont {Delbecq}, \citenamefont {Dartiailh}, \citenamefont {Cottet},\ and\ \citenamefont {Kontos}}]{viennot_out--equilibrium_2014}%
  \BibitemOpen
  \bibfield  {author} {\bibinfo {author} {\bibfnamefont {J.~J.}\ \bibnamefont {Viennot}}, \bibinfo {author} {\bibfnamefont {M.~R.}\ \bibnamefont {Delbecq}}, \bibinfo {author} {\bibfnamefont {M.~C.}\ \bibnamefont {Dartiailh}}, \bibinfo {author} {\bibfnamefont {A.}~\bibnamefont {Cottet}},\ and\ \bibinfo {author} {\bibfnamefont {T.}~\bibnamefont {Kontos}},\ }\bibfield  {title} {\bibinfo {title} {Out-of-equilibrium charge dynamics in a hybrid circuit quantum electrodynamics architecture},\ }\href {https://doi.org/10.1103/PhysRevB.89.165404} {\ \textbf {\bibinfo {volume} {89}},\ \bibinfo {pages} {165404}}\BibitemShut {NoStop}%
\bibitem [{\citenamefont {Hecker}\ \emph {et~al.}({\natexlab{b}})\citenamefont {Hecker}, \citenamefont {M{\"o}ller}, \citenamefont {Deu{\ss}en}, \citenamefont {Dulisch}, \citenamefont {Banszerus}, \citenamefont {Watanabe}, \citenamefont {Taniguchi}, \citenamefont {Volk},\ and\ \citenamefont {Stampfer}}]{hecker_probing_2026}%
  \BibitemOpen
  \bibfield  {author} {\bibinfo {author} {\bibfnamefont {K.}~\bibnamefont {Hecker}}, \bibinfo {author} {\bibfnamefont {S.}~\bibnamefont {M{\"o}ller}}, \bibinfo {author} {\bibfnamefont {T.}~\bibnamefont {Deu{\ss}en}}, \bibinfo {author} {\bibfnamefont {H.}~\bibnamefont {Dulisch}}, \bibinfo {author} {\bibfnamefont {L.}~\bibnamefont {Banszerus}}, \bibinfo {author} {\bibfnamefont {K.}~\bibnamefont {Watanabe}}, \bibinfo {author} {\bibfnamefont {T.}~\bibnamefont {Taniguchi}}, \bibinfo {author} {\bibfnamefont {C.}~\bibnamefont {Volk}},\ and\ \bibinfo {author} {\bibfnamefont {C.}~\bibnamefont {Stampfer}},\ }\href {https://doi.org/10.48550/arXiv.2605.12257} {\bibinfo {title} {Probing charge noise in bilayer graphene quantum dots by landau-zener-st{\"u}ckelberg-majorana spectroscopy}} ({\natexlab{b}}),\ \Eprint {https://arxiv.org/abs/2605.12257 [cond-mat.mes-hall]} {2605.12257 [cond-mat.mes-hall]} \BibitemShut {NoStop}%
\bibitem [{\citenamefont {Petta}\ \emph {et~al.}()\citenamefont {Petta}, \citenamefont {Johnson}, \citenamefont {Marcus}, \citenamefont {Hanson},\ and\ \citenamefont {Gossard}}]{petta_manipulation_2004}%
  \BibitemOpen
  \bibfield  {author} {\bibinfo {author} {\bibfnamefont {J.~R.}\ \bibnamefont {Petta}}, \bibinfo {author} {\bibfnamefont {A.~C.}\ \bibnamefont {Johnson}}, \bibinfo {author} {\bibfnamefont {C.~M.}\ \bibnamefont {Marcus}}, \bibinfo {author} {\bibfnamefont {M.~P.}\ \bibnamefont {Hanson}},\ and\ \bibinfo {author} {\bibfnamefont {A.~C.}\ \bibnamefont {Gossard}},\ }\bibfield  {title} {\bibinfo {title} {Manipulation of a single charge in a double quantum dot},\ }\href {https://doi.org/10.1103/PhysRevLett.93.186802} {\ \textbf {\bibinfo {volume} {93}},\ \bibinfo {pages} {186802}}\BibitemShut {NoStop}%
\bibitem [{\citenamefont {Banszerus}\ \emph {et~al.}({\natexlab{a}})\citenamefont {Banszerus}, \citenamefont {Hecker}, \citenamefont {M{\"o}ller}, \citenamefont {Icking}, \citenamefont {Watanabe}, \citenamefont {Taniguchi}, \citenamefont {Volk},\ and\ \citenamefont {Stampfer}}]{banszerus_spin_2022}%
  \BibitemOpen
  \bibfield  {author} {\bibinfo {author} {\bibfnamefont {L.}~\bibnamefont {Banszerus}}, \bibinfo {author} {\bibfnamefont {K.}~\bibnamefont {Hecker}}, \bibinfo {author} {\bibfnamefont {S.}~\bibnamefont {M{\"o}ller}}, \bibinfo {author} {\bibfnamefont {E.}~\bibnamefont {Icking}}, \bibinfo {author} {\bibfnamefont {K.}~\bibnamefont {Watanabe}}, \bibinfo {author} {\bibfnamefont {T.}~\bibnamefont {Taniguchi}}, \bibinfo {author} {\bibfnamefont {C.}~\bibnamefont {Volk}},\ and\ \bibinfo {author} {\bibfnamefont {C.}~\bibnamefont {Stampfer}},\ }\bibfield  {title} {\bibinfo {title} {Spin relaxation in a single-electron graphene quantum dot},\ }\href {https://doi.org/10.1038/s41467-022-31231-5} {\ \textbf {\bibinfo {volume} {13}},\ \bibinfo {pages} {3637} ({\natexlab{a}})}\BibitemShut {NoStop}%
\bibitem [{\citenamefont {Banszerus}\ \emph {et~al.}({\natexlab{b}})\citenamefont {Banszerus}, \citenamefont {Hecker}, \citenamefont {Wang}, \citenamefont {M{\"o}ller}, \citenamefont {Watanabe}, \citenamefont {Taniguchi}, \citenamefont {Burkard}, \citenamefont {Volk},\ and\ \citenamefont {Stampfer}}]{banszerus_phonon-limited_2025}%
  \BibitemOpen
  \bibfield  {author} {\bibinfo {author} {\bibfnamefont {L.}~\bibnamefont {Banszerus}}, \bibinfo {author} {\bibfnamefont {K.}~\bibnamefont {Hecker}}, \bibinfo {author} {\bibfnamefont {L.}~\bibnamefont {Wang}}, \bibinfo {author} {\bibfnamefont {S.}~\bibnamefont {M{\"o}ller}}, \bibinfo {author} {\bibfnamefont {K.}~\bibnamefont {Watanabe}}, \bibinfo {author} {\bibfnamefont {T.}~\bibnamefont {Taniguchi}}, \bibinfo {author} {\bibfnamefont {G.}~\bibnamefont {Burkard}}, \bibinfo {author} {\bibfnamefont {C.}~\bibnamefont {Volk}},\ and\ \bibinfo {author} {\bibfnamefont {C.}~\bibnamefont {Stampfer}},\ }\bibfield  {title} {\bibinfo {title} {Phonon-limited valley lifetimes in single-particle bilayer graphene quantum dots},\ }\href {https://doi.org/10.1103/dkgn-pfjb} {\ \textbf {\bibinfo {volume} {112}},\ \bibinfo {pages} {035409} ({\natexlab{b}})}\BibitemShut {NoStop}%
\end{thebibliography}%

\end{document}